\documentclass{aa}  
\usepackage{graphicx}
\usepackage{txfonts}
\usepackage{natbib}
\begin{document}
   \title{The X-ray activity--rotation relation of T Tauri stars in Taurus-Auriga}


   \author{K.~R.~Briggs
          \inst{1}
          \and
          M.~G\"udel\inst{1}
	  \and
	  A.~Telleschi\inst{1}
          \and
	  T.~Preibisch\inst{2}
	  \and
	  B.~Stelzer\inst{3}
	  \and 
	  J.~Bouvier\inst{4}
	  \and
	  L.~Rebull\inst{5}
          \and
	  M.~Audard\inst{6,7,8}
	  \and
	  L.~Scelsi\inst{9}
          \and
	  G.~Micela\inst{3}
          \and
          N.~Grosso\inst{4}
	  \and
          F.~Palla\inst{10}
          }

   \offprints{K.~R.~Briggs}

   \institute{Paul Scherrer Institut, Villigen und W\"urenlingen,
              CH-5232, Switzerland\\
              \email{briggs@astro.phys.ethz.ch}
	\and
             Max-Planck-Institut f\"ur Radioastronomie, 
             Auf dem H\"ugel 69, 53121 Bonn, Germany
	\and
             INAF, Osservatorio Astronomico di Palermo, 
             Piazza del Parlamento 1, 90134 Palermo, Italy      
	\and
             Laboratoire d'Astrophysique de Grenoble, 
	     Universit\'e Joseph Fourier, BP 53, 38041, Grenoble Cedex 9, 
             France
        \and
             Spitzer Science Center, Caltech M/S 220-6, 
             1200 East California Boulevard, Pasadena, CA 91125, USA
        \and
             Columbia Astrophysics Laboratory, Mail Code 5247, 
             550 West 120th Steet, New York, NY 10027, USA
        \and
             Integral Science Data Center, Ch. d'Ecogia 16, CH-1290 Versoix, 
             Switzerland
        \and
             Geneva Observatory, University of Geneva, Ch. des Maillettes 51,
             1290 Sauverny, Switzerland
        \and
             Dipartimento di Scienze Fisiche ed Astronomiche, 
             Sezione di Astronomia, Universit\`a di Palermo, 
             Piazza del Parlamento 1, 90134 Palermo, Italy
        \and
             INAF, Osservatorio Astrofisica di Arcetri, Largo E. Fermi 5,
             50125 Florence, Italy
             }

   \date{Received 28th November, 2006; accepted 10th January, 2007}

 
  \abstract
{The Taurus-Auriga star-forming complex hosts the only population of T~Tauri stars in which an anticorrelation of X-ray activity and rotation period has been observed.
}
{We aim to explain the origin of the X-ray activity--rotation relation in Taurus-Auriga. We also aim to put the X-ray activity of these stars into the context of the activity of late-type main-sequence stars and T~Tauri stars in the Orion Nebula Cluster.}
{We have used {\it XMM-Newton}'s European Photon Imaging Cameras to perform the most sensitive survey to date of X-ray emission (0.3-10~keV) from young stars in Taurus-Auriga. We investigated the dependences of X-ray activity measures -- X-ray luminosity, $L_{\rm X}$, its ratio with the stellar luminosity, $L_{\rm X}/L_{\star}$, and the surface-averaged X-ray flux, $F_{\rm XS}$ -- on rotation period and compared them with predictions based solely on the observed dependence of $L_{\rm X}$ on a star's $L_{\star}$ and whether it is accreting or not. We tested for differences in the distributions of $L_{\rm X}/L_{\star}$ of fast and slow rotators, accretors and non-accretors, and compared the dependence of $L_{\rm X}/L_{\star}$ on the ratio of the rotation period and the convective turnover timescale, the Rossby number, with that of late-type main-sequence stars.}
{We found significant anticorrelations of $L_{\rm X}$ and $F_{\rm XS}$ with rotation period, but these could be explained by the typically higher stellar luminosity and effective temperature of fast-rotators in Taurus-Auriga and a near-linear dependence of $L_{\rm X}$ on $L_{\star}$. We found no evidence for a dependence of $L_{\rm X}/L_{\star}$ on rotation period, but for accretors to have lower $L_{\rm X}/L_{\star}$ than non-accretors at all rotation periods. The Rossby numbers of accretors and non-accretors were found to be the same as those of late-type main-sequence stars showing saturated X-ray emission.}
{Non-accreting T~Tauri stars show X-ray activity entirely consistent with the saturated activity of fast-rotating late-type main-sequence stars. Accreting T~Tauri stars show lower X-ray activity, but this cannot be attributed to their slower rotation.}


\authorrunning{K. R. Briggs}
\titlerunning{The X-ray activity--rotation relation in Taurus-Auriga}

   \maketitle 
%

\section{Introduction}
\label{sec-intro}

T~Tauri stars are young stars contracting toward the main sequence, a subclass of which, the `classical' T~Tauri stars, show signatures that they are actively accreting material from circumstellar accretion disks.
Both accreting and non-accreting T~Tauri stars exhibit strong, variable, X-ray emission which provides evidence for hot coronae generated by magnetic activity. Key open questions concerning this emission are the extent to which this magnetic activity is analogous to that exhibited by the Sun and similar stars, and what effect interaction with the circumstellar material, particularly through accretion, has on the magnetic activity.

In the Sun, and all stars whose structure is composed of a radiative interior and a convective envelope, magnetic activity is believed to be predominantly generated by a dynamo that is located in a shell at the interface of the radiative and convective zones and is driven by (differential) rotational and convective motions \citep[an $\alpha\Omega$ dynamo:][]{parker}. This is supported by observations of signatures of magnetic activity, such as chromospheric H$\alpha$ or Ca~H and K line emission or coronal X-ray emission, in stars with convective envelopes. The strengths of these activity signatures correlate directly with the projected rotation velocity, $v \sin i$, and inversely with the rotation period of the star, $P_{\rm rot}$ \citep[e.g.][]{pallavicini}. They furthermore show a tighter inverse correlation with the Rossby number, $R_0$, which is the ratio of the rotation period to the convective turnover timescale at the base of the convective envelope, $\tau_{\rm conv}$ \citep[e.g.][]{noyes}.

The strengths of the activity signatures are observed to saturate at fast rotation velocities, short rotation periods and low Rossby numbers \citep{vilhu84,vw87}. It is not yet known whether this is due to a saturation of the dynamo itself or constraints on the available volume of the corona \citep[see e.g.][]{ju99}.

\citet{pizzolato} have shown that the ratio of the X-ray luminosity to the stellar bolometric luminosity, $L_{\rm X}/L_{\star}$, for main-sequence stars of spectral types from G to early M is well-determined by the Rossby number, $R_0$, such that $L_{\rm X}/L_{\star} \approx 10^{-3.2} (R_0/R_{0,{\rm sat}})^{-2}$ for $R_0 > R_{0,{\rm sat}}$ and $L_{\rm X}/L_{\star} \approx 10^{-3.2}$ for $R_0 < R_{0,{\rm sat}}$, where $R_{0,{\rm sat}} \approx 0.2$ is the Rossby number at which saturation sets in.

T~Tauri stars with masses greater than approximately 0.3~$M_{\odot}$ are expected to be initially fully-convective and to develop a solar-like structure after nuclear fusion begins in the core. Stars with masses less than approximately 0.3~$M_{\odot}$ are expected to remain fully-convective. The majority of T~Tauri stars are thus expected to have long convective turnover timescales of 100--300~d. Despite the fact that an $\alpha\Omega$ dynamo cannot operate in fully-convective main-sequence stars, there is no empirical evidence for different characteristics in the magnetic activity of these stars: fast rotators show activity at the saturated level and only slow rotators show activity well below the saturated level \citep{del98,mb03}. 

The T~Tauri stars in Taurus-Auriga stand out as the only population of pre-main sequence stars in which an anticorrelation of X-ray activity with rotation period has been observed \citep{bou90,dam95,neu95,sn01}. 
The largest survey to date of the X-ray emission from T~Tauri stars, the {\it Chandra} Ultradeep Orion Project (COUP) observation of almost 600 T~Tauri stars in the Orion Nebula Cluster (ONC), found no such anticorrelation \citep{coup}. The Rossby numbers of all the COUP stars with measured rotation periods placed them in the saturated regime of main-sequence stars, and their $L_{\rm X}/L_{\star}$ was largely consistent with the saturation level of main-sequence stars but with a very large scatter. A strong dependence of X-ray luminosity on the stellar bolometric luminosity was found, as had been seen in previous surveys of X-ray emission from T~Tauri stars \citep[e.g.][]{ic348}. Accreting T~Tauri stars were found to have generally lower X-ray activity levels than non-accretors.

The lower X-ray activity of accreting T~Tauri stars has also been observed in Taurus-Auriga \citep{dam95,neu95,sn01}, and has been usually attributed to their typically slower rotation and an anticorrelation of X-ray activity with rotation period.

We use the results of a new X-ray survey of Taurus-Auriga, the {\it XMM-Newton} Extended Survey of the Taurus Molecular Cloud (XEST), described in Sect.~\ref{sec-xest}, to reinvestigate the dependence of X-ray emission on rotation in this star-forming region. Our sample selection and data analysis methods are described in Sects.~\ref{sec-sample} and \ref{sec-data}, respectively.
In Sect.~\ref{sec-act-rot} we present, and propose an explanation for, the observed dependences of X-ray activity measures on rotation period. In Sect.~\ref{sec-lxls}, we explore the role of rotation in the different activity levels of accreting and non-accreting stars.
We investigate the dependence of X-ray activity on Rossby number in Sect.~\ref{sec-rossby}, to examine the observed activity in the context of the activity of solar-like main-sequence stars and of T~Tauri stars in the ONC studied by COUP.
In Sect.~\ref{sec-rosat}, we make a careful comparison with, and reexamination of, the most comprehensive previous X-ray study of Taurus-Auriga, made by \citet{sn01} with the {\it ROSAT} Position Sensitive Proportional Counter. 
In Sect.~\ref{sec-summary}, we conclude by summarizing our findings and identifying outstanding questions and suggesting prospective observing strategies which could address them.

\section{The {\it XMM-Newton} Extended Survey of the Taurus Molecular Cloud (XEST)}
\label{sec-xest}

The XEST comprises 28 observations with exposure times of at least 30~ks of regions of the Taurus-Auriga complex that are most densely-populated by T Tauri stars. {\it XMM-Newton}'s coaligned European Photon Imaging Camera (EPIC) PN, MOS1 and MOS2 detectors each have a full field of view approximately 30 arcmin in diameter and perform simultaneous imaging, timing and spectroscopy of X-rays in the energy range 0.2--12~keV, with respective resolutions of 6~arcsec (FWHM), less than 2.5~s, and 50--200~eV ($E/\Delta E \approx 20$--50).
 The survey observations and data reduction procedures are described in detail in \citet{xest-overview}. 

XEST makes key improvements over previous surveys of Taurus-Auriga using the {\it Einstein} Imaging Proportional Camera (IPC) and {\it ROSAT} Position Sensitive Proportional Camera (PSPC) \citep{bou90,dam95,neu95,sn01} thanks primarily to the greater sensitivity, broader energy band and higher spectral resolution of the EPIC detectors, augmented by longer exposure times. XEST reaches X-ray luminosities of $10^{28}$\,erg\,s$^{-1}$ at the 140~pc distance of Taurus-Auriga \citep{tmc-dist}, approximately an order of magnitude lower than the {\it ROSAT} PSPC pointing survey \citep{sn01}, enabling the detection of almost all known T Tauri stars in the survey area down to the substellar mass limit for the first time. This is vital for accurately quantifying the dependence of X-ray activity on parameters such as stellar luminosity. T Tauri stars, particularly accreting T~Tauri stars, are typically surrounded by circumstellar material or observed through molecular cloud material that absorbs soft X-ray emission from the star. The broad bandpass of EPIC allows it to detect the less absorbed harder emission. The EPIC spectra enabled the absorption to be measured for each T Tauri star and accounted for in calculating its X-ray luminosity. This is crucial for accurate determination of X-ray luminosities and was not generally possible with the data obtained by the {\it Einstein} and {\it ROSAT} surveys. 
Additionally, since the {\it ROSAT} surveys many more low-mass members of Taurus-Auriga have been identified by optical and near-infrared photometric surveys \citep[e.g.][]{luhman03}, making the known population of T Tauri stars much more complete.

\section{Sample selection}
\label{sec-sample}

We performed a literature study to compile a catalogue of recognised members of Taurus-Auriga and their relevant data, such as stellar luminosity, $L_{\star}$, effective temperature, $T_{\rm eff}$, multiplicity, photometric rotation period, $P_{\rm rot}$, projected equatorial rotation velocity, $v \sin i$, and signatures of accretion such as the equivalent width of the H$\alpha$ line. This catalogue and the many sources of information used in its compilation are found, along with the XEST X-ray data for these stars, in \citet{xest-overview}.

The XEST data provide an almost detection-complete sample of more than 100 T Tauri stars (excluding Class~I protostars, Herbig~Ae stars and substellar objects) evenly divided between accretors and non-accretors and well-spread in masses from 0.1 to 2.5\,M$_{\odot}$. We have chosen to restrict our study to T~Tauri stars of spectral types M3 and earlier ($T_{\rm eff} > 3400$~K). No star in the XEST sample of later spectral type has a measured photometric period, and few such stars have measured $v \sin i$. Stars of later spectral type are typically less-well studied and are not expected to develop a solar-like structure.

We have excluded four undetected accreting T Tauri stars for which $L_{\rm X}$ upper limits could not be calculated due to lack of knowledge of absorption (\object{Coku-Tau~1}, \object{HH~30}, \object{ITG~33A} and \object{IRAS S04301+261}), two T~Tauri stars whose characteristic X-ray activity could not be assessed because a decay from a large flare persisted throughout their XEST observation (\object{DH~Tau} and \object{V830~Tau}), and all T~Tauri stars for which $T_{\rm eff}$ and/or $L_{\star}$ were not available, or whose $T_{\rm eff}$ and $L_{\star}$ placed them outside the stellar evolutionary model grids of \citet{siess}. The decay of a large flare also dominated the XEST observation of \object{FS~Tau}, but we have used additional {\it Chandra} data to assess its characteristic X-ray activity.

Our sample contains a total of 74 T~Tauri stars, 47 accretors and 27 non-accretors. These include 23 stars with a measured photometric rotation period, of which 13 are accretors and 10 are non-accretors. There are 47 stars that have a measured $v \sin i$, of which 30 are accretors and 17 are non-accretors, and this sample includes all stars with a measured photometric rotation period except for the accretor \object{XZ~Tau}.

\section{Data Analysis}
\label{sec-data}

Our study comprises four investigations. The data analysis performed in these investigations is described in this section, while the results of each investigation are presented and discussed in a dedicated section in Sects.~\ref{sec-act-rot}--\ref{sec-rosat}.

\begin{figure}
\centering
\includegraphics[angle=270,width=0.45\textwidth]{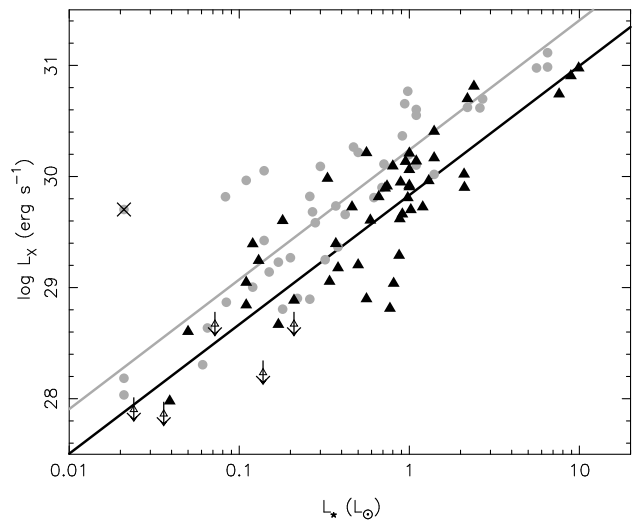}
\caption{The dependence of X-ray luminosity upon stellar luminosity for the accreting (black triangles) and non-accreting (grey circles) T~Tauri stars in Taurus-Auriga observed by {\it XMM-Newton}. The black and grey lines show the best-fitting correlations for accretors and non-accretors, respectively, found using the EM algorithm in ASURV \citep{xest-acc}. Open symbols and downward-pointing arrows mark upper limits. The cross shows \object{KPNO-Tau~8}, which was excluded from the fitting.}
\label{fig-lx-ls}
\end{figure}

Firstly, we examined the correlations with rotation period of the three most commonly-used measures of X-ray activity: the X-ray luminosity, $L_{\rm X}$, its ratio with the stellar bolometric luminosity, $L_{\rm X}/L_{\star}$, and the surface-averaged X-ray flux, $F_{\rm XS} = L_{\rm X}/4 \pi R_{\star}^2$. In order to compare to previous studies we have treated accreting and non-accreting stars as a single sample. We tried to understand the observed activity--rotation relations in terms of the near-linear dependency of $L_{\rm X}$ on $L_{\star}$, and the typically lower $L_{\rm X}$ of accreting stars, which have been consistently observed in populations of T~Tauri stars that do not show an anticorrelation of activity on rotation \citep[e.g.][]{coup,ic348}. We used the correlations of $L_{\rm X}$ on $L_{\star}$ derived separately for accretors and non-accretors in the XEST study by \citet[see Fig.~\ref{fig-lx-ls}]{xest-acc} to calculate the expected $L_{\rm X}$, and hence the $L_{\rm X}/L_{\star}$ and $F_{\rm XS}$, of each star in our sample based on its $L_{\star}$ and whether it was accreting or not. We compared the resulting correlations with rotation period of the expected and the observed activity measures. The results are presented and discussed in Sect.~\ref{sec-act-rot}.

Secondly, we tested the hypothesis that the lower $L_{\rm X}/L_{\star}$ of accretors compared to non-accretors is due to their slower rotation period. In this analysis we have divided the full sample described in Sect.~3, into subsamples based on rotation period calculated from $v \sin i$ and on whether the star was accreting or not and statistically compared the distributions of $L_{\rm X}/L_{\star}$ of the subsamples. The results are presented and discussed in Sect.~\ref{sec-lxls}.

Thirdly, we compared the X-ray activity of the T~Tauri stars in our sample with the activity of solar-like main-sequence stars and T~Tauri stars in the ONC based on the dependence of $L_{\rm X}/L_{\star}$ on Rossby number. The results are presented and discussed in Sect.~\ref{sec-rossby}.

Finally we compared our activity--rotation relations with those obtained by the previous most-comprehensive X-ray survey of Taurus-Auriga, made by \citet{sn01} using the {\it ROSAT} PSPC. The results of this comparison are presented and discussed in Sect.~\ref{sec-rosat}.

The X-ray activity measures and rotation periods we used are described in Sects.~\ref{sec-lx} and \ref{sec-rot}, respectively. The criterion we used to distinguish accretors and non-accretors is given in Sect.~\ref{sec-acc}. The estimation of the convective turnover timescale for each of the stars in our sample is described in Sect.~\ref{sec-tconv}. The statistical methods used in our investigations are described in Sect.~\ref{sec-stats}.

\subsection{Determination of the X-ray activity measures}
\label{sec-lx}

X-ray activity measures, $L_{\rm X}$, $L_{\rm X}/L_{\star}$ and $F_{\rm XS}$, were calculated for each star using the data tabulated in \citet{xest-overview}.The $L_{\rm X}$ of each observed T~Tauri star was calculated in \citet{xest-overview} as follows. A spectral fit of each source was made in XSPEC \citep{xspec} using a model composed of an optically-thin collisionally-ionized, so-called `coronal', plasma \citep[APEC;][]{apec}, having a two-power-law distribution of emission measure with temperature, multiplied with a photoelectric absorption model (WABS). The emission measure distribution was cut off at $10^6$ and $10^8$~K. The elemental abundances of the plasma were fixed to values derived from high-resolution X-ray spectra of young stars\footnote{The abundance values used were, with respect to the solar photospheric abundances of \citet{ag89}: C=0.45, N=0.788, O=0.426, Ne=0.832, Mg=0.263, Al=0.5, Si=0.309, S=0.417, Ar=0.55, Ca=0.195, Fe=0.195, Ni=0.195.}. The power-law index at low energies, $\alpha$, is unconstrained in EPIC spectra and was fixed to $+2$, as observed in young solar analogues \citep{solar-an}. The free parameters were the break temperature of the two power laws, $6.3 \ge \log T_0 \ge 7.5$, the power-law index at high energies, $-3 \ge \beta \ge +1$, the normalization, and the equivalent hydrogen column density of the absorbing gas, $N_{\rm H}$. We integrated the flux under the best-fitting model between 0.3 and 10.0~keV and assumed a distance of 140~pc to calculate the X-ray luminosity. 
Time intervals in which a star was clearly flaring were removed before spectral analysis if possible.

We excluded observations of stars in which the emission was clearly dominated by a flare throughout the observation. If there were two or more acceptable observations of the same star we calculated the X-ray luminosity as the logarithmic average of the individual $L_{\rm X}$ determinations.

For each observation of a star we estimated the uncertainty in $L_{\rm X}$ by finding the 68 per cent confidence interval in $N_{\rm H}$ and calculating $L_{\rm X}$ for the best-fitting models with $N_{\rm H}$ fixed to the lower and upper limits of this confidence interval.
$L_{\rm X}$ was poorly constrained when there were few counts, particularly when $N_{\rm H}$ was appreciable ($>$ few $\times 10^{21}$~cm$^{-2}$) and the slope of the spectrum at energies above 2~keV was not well-defined. A degeneracy between $\log T_0$ and $N_{\rm H}$ sometimes developed, that in the worst case led to well-fitting models with very low $\log T_0$ and very high $N_{\rm H}$, hence high $L_{\rm X}$. Such cool temperatures were not observed in good-quality spectra showing low $N_{\rm H}$ and it is therefore likely that such models highly overestimate $L_{\rm X}$. In such cases the error bars plotted throughout this work extend from these probably unrealistically high values of $L_{\rm X}$ to lower values corresponding to more realistic hotter models with lower $N_{\rm H}$.
In some poorly constrained fits we fixed $\beta$, usually the least well-constrained parameter, to a typical value of $-1$. In \citet{xest-overview} we also fixed $N_{\rm H}$ in several particularly poorly-constrained cases to the value indicated by the visual or near-infrared extinction. However, this results in unrepresentative uncertainies and we have reanalysed those four cases here with just $\beta$ fixed to $-1$. The results of this additional analysis are given in Table~\ref{tab-newspecfits}.

We have also performed additional spectral analysis of four jet-driving accreting T~Tauri stars, \object{DG~Tau~A}, \object{GV~Tau}, \object{DP~Tau} and \object{CW~Tau}, whose X-ray spectra could not be well-described by a model with just a single absorbing column. \citet{xest-jets} found instead a little-absorbed unvarying low-temperature component, which they proposed to be due to shocks in the jets, and a highly-absorbed and variable hot component, which they attributed to a corona. We have modelled the hot component with the spectral model described above -- rather than the absorbed isothermal plasma used by \citet{xest-jets} -- and, the cool component with an isothermal model with photoelectric absorption. The hot components of \object{DG~Tau~A} and \object{GV~Tau} were strongly variable in their XEST observations. In these cases we have first used the average spectrum from the whole observation to determine the best-fitting parameters of the unvarying cool component. The parameters of the cool component were then fixed to these values when we fitted the spectrum of the time intervals when the star appeared to be in a low or `quiescent' X-ray emission state to derive characteristic parameters for the hot component. \object{CW~Tau} required no cool component when we excluded energies below 0.5~keV. We fixed $\beta$ to $-1$ in these fits. Nevertheless, $\log T_0$ was unconstrained in all cases, so we have calculated $L_{\rm X}$ for $\log T_0 = 7.0$ and have expressed the range of uncertainties found when $\log T_0$ was a free parameter (see Table~\ref{tab-newspecfits}).

Just one star in our sample, \object{FV~Tau/c}, was not detected in the XEST observation. \object{FV~Tau/c} has no measured photometric rotation period or $v \sin i$. As described in \citet{xest-overview}, we calculated a 95 per cent confidence upper limit to its EPIC count-rate and used a spectrum with $\log T_0 = 7.0$ and $\beta = -1$, with $N_{\rm H}$ estimated from its optical extinction, to calculate an upper limit to its X-ray luminosity.

\begin{table*}
\centering
\caption{The results of spectral fits additional to those performed in \citet{xest-overview}.}
\begin{tabular}{llcccccr}
\hline
XEST & Name & $N_{\rm H}$ ($1\sigma$ range) & $\log T_0$ ($1\sigma$ range) & $L_{\rm X}$ ($1\sigma$ range) & $\log (L_{\rm X}/L_{\star})$ & $\chi^2_{\rm red}$ & dof\\
 & & $10^{22}$\,cm$^{-2}$ & K & $10^{30}$\,erg\,s$^{-1}$ & & & \\
\hline
04-010 & \object{GH~Tau~AB} & 0.48 \small{(0.33,0.63)} & 6.4 \small{(6.3,7.3)} & 0.39 \small{(0.29,0.50)} & $-3.91$ & 0.89 & 4\\
09-010 & \object{HO~Tau~AB} & 0.27 \small{(0.09,0.44)} & 6.6 \small{(6.3,6.9)} & 0.064 \small{(0.034,0.13)} & $-4.00$ & 1.11 & 3\\
07-011 & \object{JH~223}   & 0.19 \small{(0.14,0.31)} & 6.6 \small{(6.3,6.7)} & 0.099 \small{(0.078,0.18)} & $-3.84$ & 0.89 & 4\\
11-079 & \object{CFHT-Tau~21} & 1.72 \small{(1.20,2.20)} & 7.2 \small{(6.3,7.5)} & 0.22 \small{(0.15,0.36)} & $-3.82$ & 0.68 & 7\\
02-022 & \object{DG~Tau~A} & 4.50 \small{(2.07,7.61)} & 7.0 \small{(6.3,7.5)} & 0.55 \small{(0.17,2.39)} & $-4.08$ & 0.77 & 13\\
20-046 & \object{CW~Tau}   & 2.93 \small{(1.33,4.72)} & 7.0 \small{(6.3,7.5)} & 0.11 \small{(0.038,0.31)} & $-4.60$ & 0.94 & 9\\
10-045 & \object{DP~Tau}   & 6.31 \small{(2.98,9.02)} & 7.0 \small{(6.3,7.5)} & 0.33 \small{(0.089,0.74)} & $-3.37$ & 1.16 & 8\\
13-004 & \object{GV~Tau~A} & 3.02 \small{(1.31,5.16)} & 7.0 \small{(6.3,7.5)} & 0.55 \small{(0.19,2.86)} & $-4.11$ & 0.82 & 6\\
\hline
\end{tabular}
\label{tab-newspecfits}
\end{table*}

For recognised multiple systems not spatially resolved into their individual components by {\it XMM-Newton}, we have assumed that the $L_{\rm X}$ of the components scales linearly with $L_{\star}$, as motivated by Fig.~\ref{fig-lx-ls}. Thus, each component of a multiple system is assumed to have the same $L_{\rm X}/L_{\star}$, which is the ratio of the observed $L_{\rm X}$ and the total stellar luminosity of the system.
The multiplicity-corrected $L_{\rm X}$ of the primary is therefore the observed $L_{\rm X}$ multiplied by the fraction of the summed stellar luminosities that is contributed by the primary. This fraction was estimated using the luminosities of the individual components if these had been determined, else the magnitude differences in the $K$-band or $I$-band, as available\footnote{The $K$-band is not ideal as accretion disks can emit substantially in this band, and lower-mass companions emit a higher proportion of their bolometric flux in this band relative to higher-mass primaries. For this latter reason, however, many multiplicity studies have been performed in this band.}. 
$L_{\rm X}/L_{\star}$ was always calculated using the uncorrected observed $L_{\rm X}$ and the total stellar luminosity of the system. $F_{\rm XS} = L_{\rm X}/4 \pi R_{\star}^2 = (L_{\rm X}/L_{\star}) \sigma T_{\rm eff}^4$ was always calculated using $L_{\rm X}/L_{\star}$ and the effective temperature, $T_{\rm eff}$ of the primary.

\subsection{Rotation periods}
\label{sec-rot}

Rotational data in the form of rotation periods and spectroscopic $v \sin i$ have been taken from \citet{rebull}, where a comprehensive list of references can be found, with a small number of additions from L.~Rebull (priv. comm.), and were always assigned to the primary in the cases of unresolved multiple systems. 
For the larger sample of T~Tauri stars with measured $v \sin i$, we estimated rotation periods as $P_{\rm rot}/ \sin i = 2 \pi R_{\star} / v \sin i$.
 These are essentially upper limits to the rotation period\footnote{Two of the slowly-rotating non-accretors, \object{HBC~358} and \object{HBC~359}, have only upper limits to $v \sin i$ of 10~km\,s$^{-1}$ and hence lower limits to $P_{\rm rot}/ \sin i$ of $> 6.98$ and $> 6.72$~d, respectively.} but we note that for a randomly oriented sample $\langle \sin i \rangle = \pi/4 \approx 0.785$, the $1 \sigma$ lower limit to $\sin i$ is 0.73 and there is just a 10 per cent chance that $\sin i$ is less than 0.44. \object{RY~Tau}, \object{LkCa~21} and \object{CW~Tau} have tabulated photometric periods in \citet{xest-overview} but their high measured $v \sin i$ indicate that the photometric periods are too long to be the rotation periods \citep{bou93,bou95} and we use only the $v \sin i$ values for these three stars in this work.

\subsection{Accretion criterion}
\label{sec-acc}

Accreting pre-main sequence stars, classical T~Tauri stars, have been traditionally identified through their strong $H\alpha$ emission, believed to arise from heated accreting material. $H\alpha$ emission is also produced by chromospheric activity, but, outside of exceptional flares, is restricted by the saturation of magnetic activity and one can use the criterion $\log (L_{{\rm H}\alpha}/L_{\star}) > -3.3$ to identify accreting stars \citep{cttsdef}. It is simpler to measure the equivalent width of the $H\alpha$ line, $EW_{{\rm H}\alpha}$, but cooler stars have less continuum emission at the $H\alpha$ line and so the criterion becomes a function of spectral type. We have adopted the dependence on spectral type of the $EW_{{\rm H}\alpha}$ criterion that was proposed by \citet{cttsdef}\footnote{The criterion proposed by \citet{cttsdef2} gives identical classifications for the present sample.}, directly based on the criterion $\log (L_{{\rm H}\alpha}/L_{\star}) > -3.3$. The classification of the primary was used in cases of multiple systems not resolved by {\it XMM-Newton}.

\subsection{Determination of convective turnover timescales}
\label{sec-tconv}

\begin{figure}
\centering
\includegraphics[angle=270,width=0.45\textwidth]{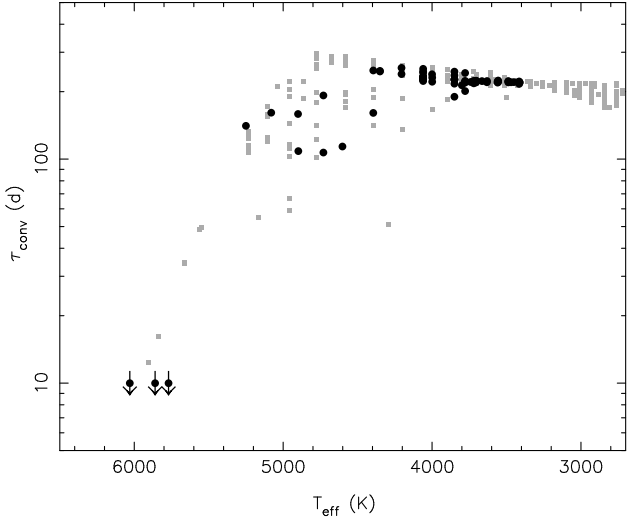}
\caption{Convective turnover timescales of T~Tauri stars in Taurus-Auriga observed by {\it XMM-Newton}, plotted as a function of effective temperature. The values have been interpolated from values calculated for T~Tauri stars in the Orion Nebula Cluster \citep[shown by grey squares, ][Preibisch, priv. comm.]{coup}, based on stellar luminosity and effective temperature.}
\label{fig-tconv}
\end{figure}

Convective turnover timescales must be determined empirically or derived from stellar models. In \citet{coup}, a convective turnover timescale was derived from a full stellar model for each of 596 T~Tauri stars in the ONC. We have determined the value of $\tau_{\rm conv}$ for each T~Tauri star in the XEST sample from the values calculated for the ONC stars (Th.~Preibisch, priv. comm.), based on its stellar luminosity and effective temperature (see Fig.~\ref{fig-tconv}). For multiple systems, the $L_{\star}$ and $T_{\rm eff}$ of the primary were used where available. The stars with $T_{\rm eff} < 4100$~K are essentially all fully-convective according to the evolutionary models of \citet{siess} and have very similar $\tau_{\rm conv}$ of 200--250~d. The stars in our sample with $T_{\rm eff}$ in the range 4100--5400~K have radiative interiors and shorter $\tau_{\rm conv}$, but these are still longer than 100~d. $\tau_{\rm conv}$ falls steeply with further increases in $T_{\rm eff}$. As the three G-type stars in the XEST sample (\object{SU~Aur}, \object{HD~283572} and \object{HP~Tau/G2}) are much less luminous than stars with similar $T_{\rm eff}$ in the ONC sample, all we can say is that the $\tau_{\rm conv}$ of these stars are shorter than the shortest value calculated for ONC stars, 11.6~d.


\begin{figure*}[!h]
\centering
\hbox{
\includegraphics[angle=270,width=0.45\textwidth]{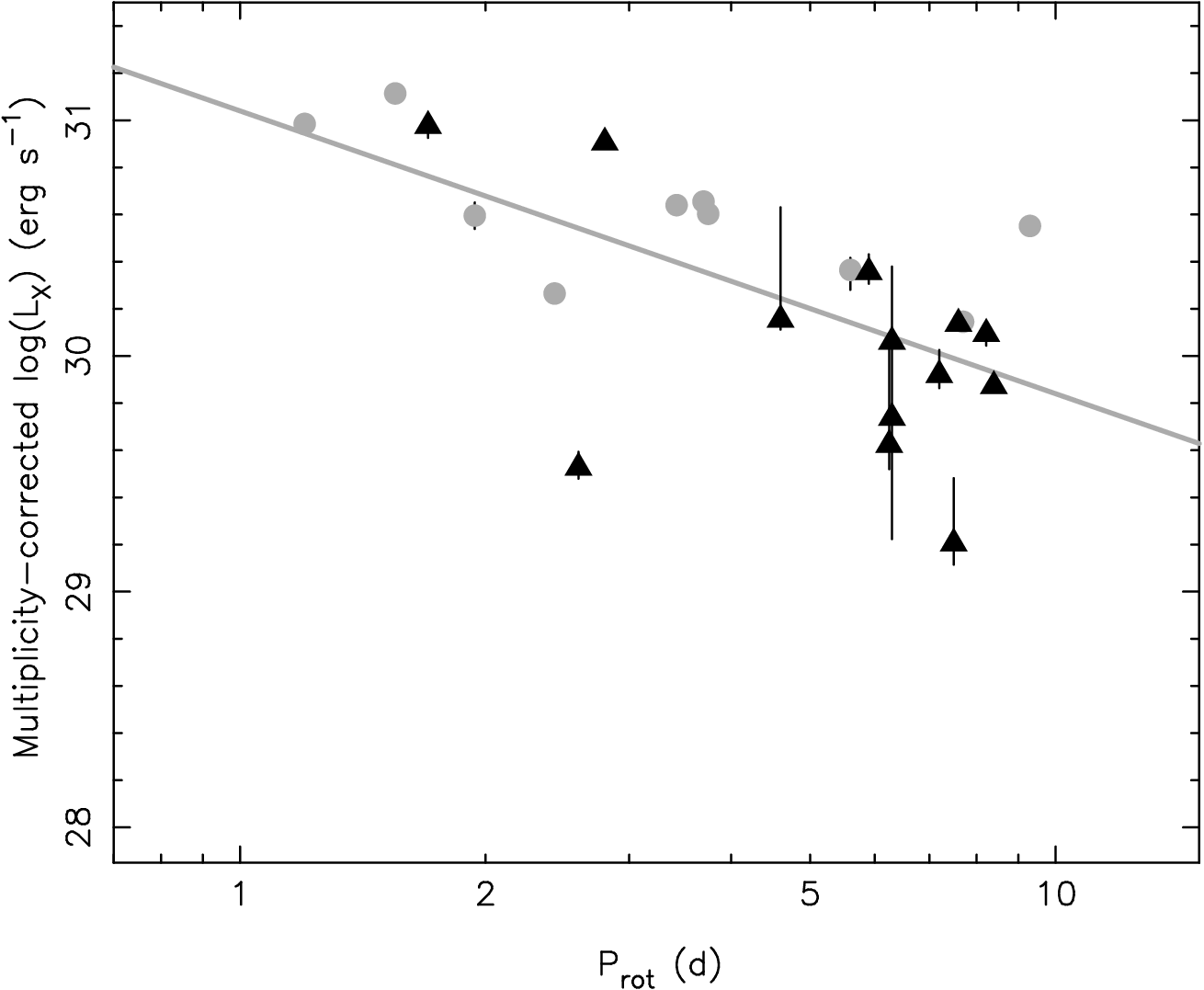}
\includegraphics[angle=270,width=0.45\textwidth]{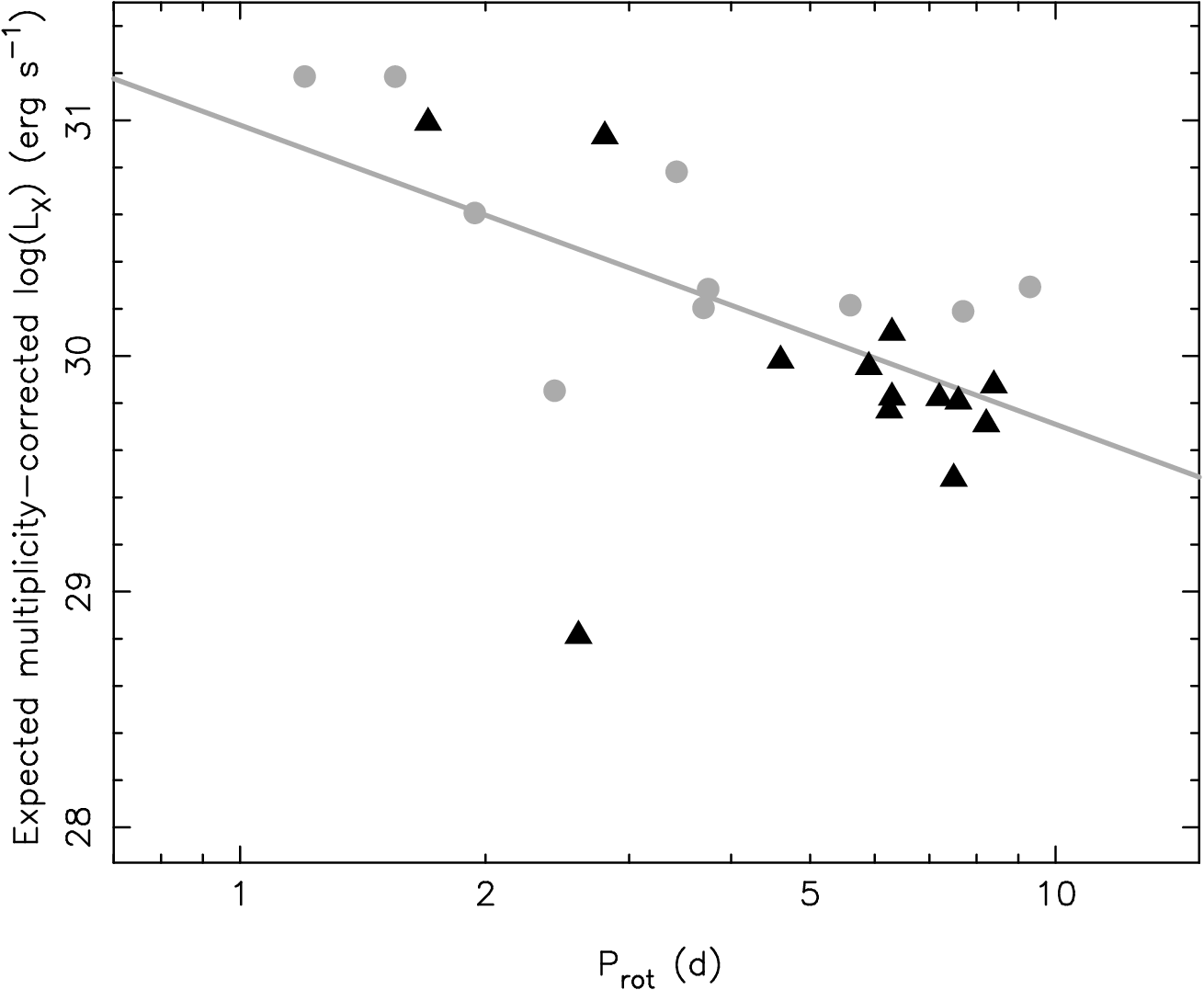}
}
\hbox{
\includegraphics[angle=270,width=0.45\textwidth]{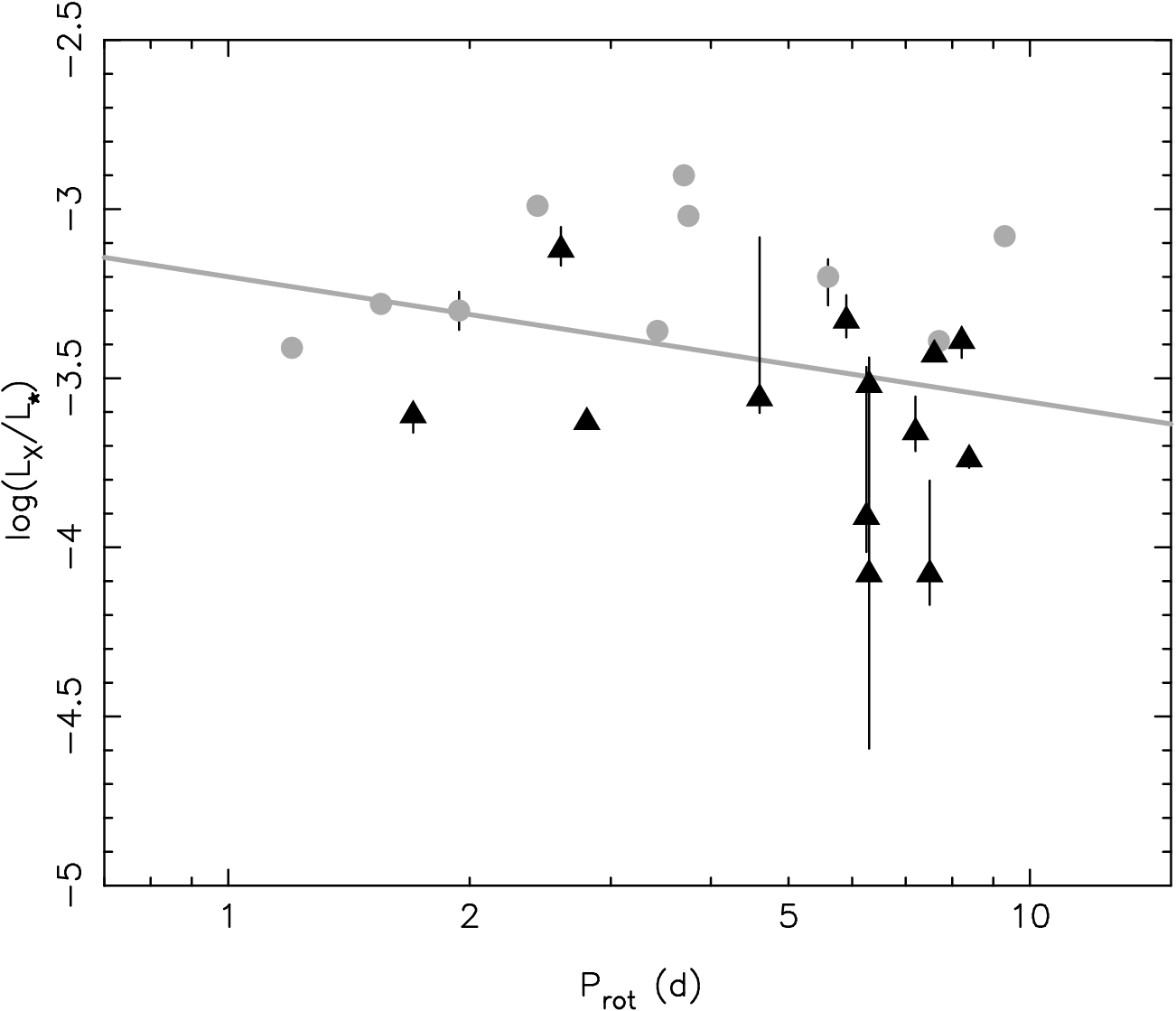}
\includegraphics[angle=270,width=0.45\textwidth]{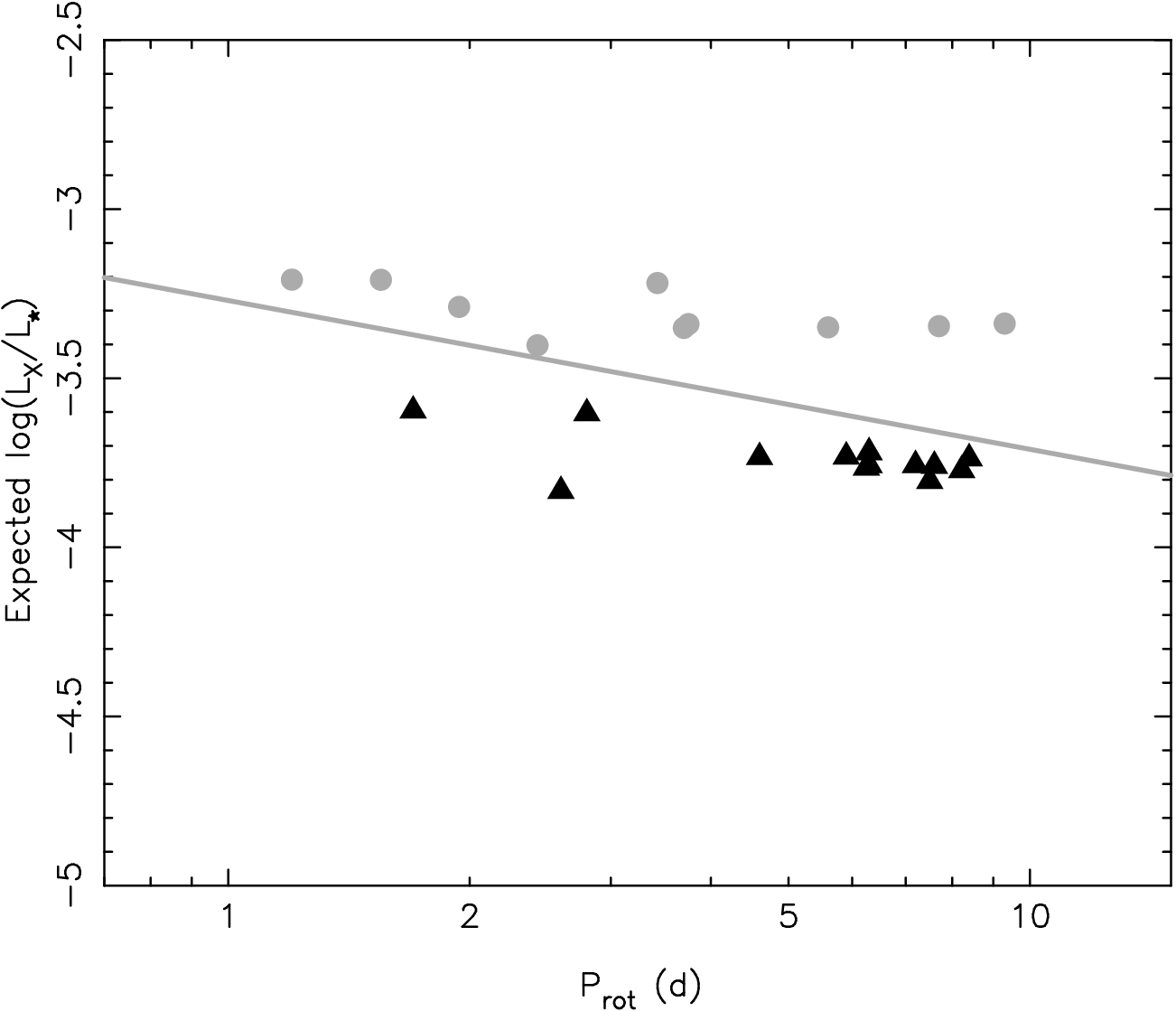}
}
\hbox{
\includegraphics[angle=270,width=0.45\textwidth]{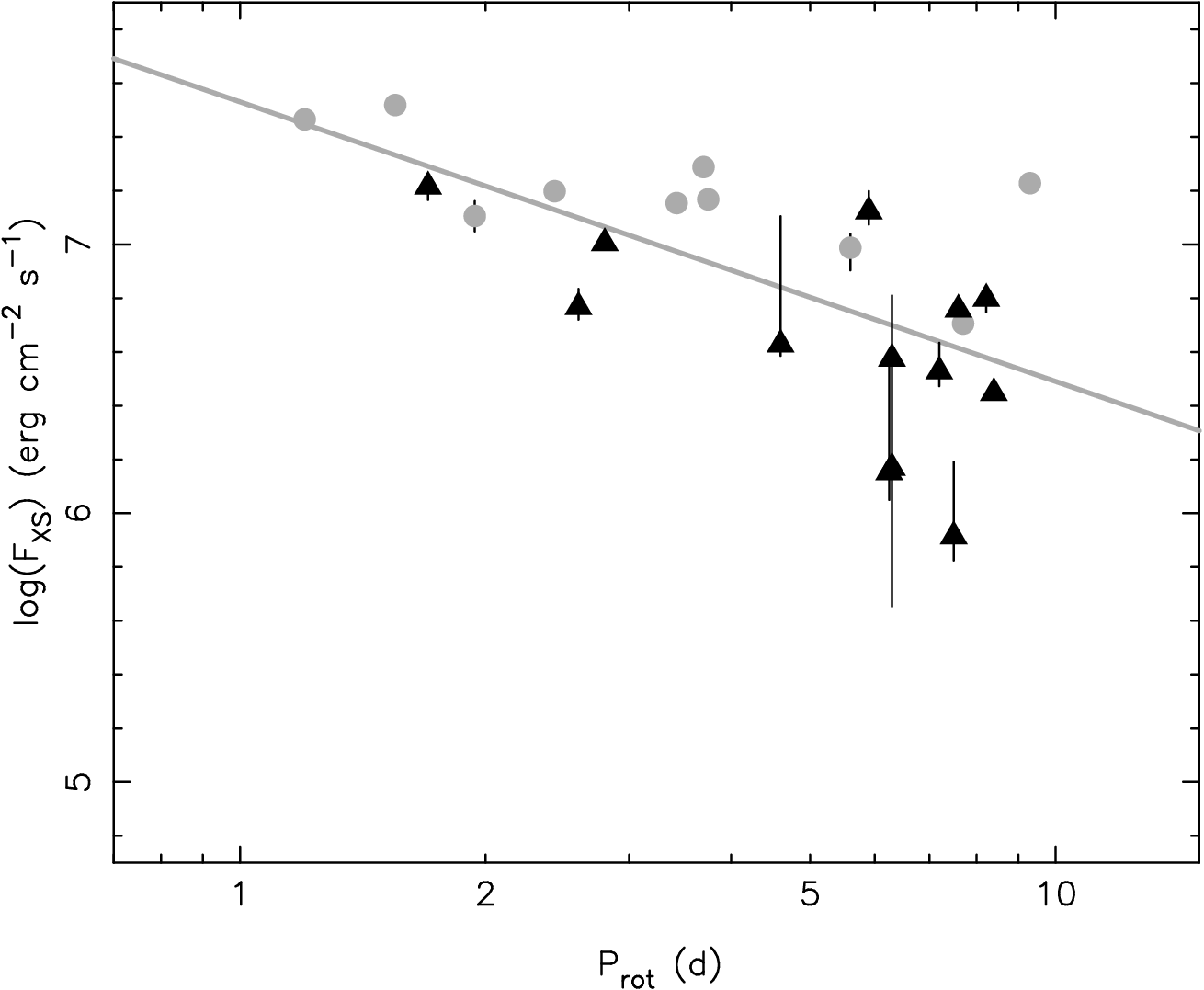}
\includegraphics[angle=270,width=0.45\textwidth]{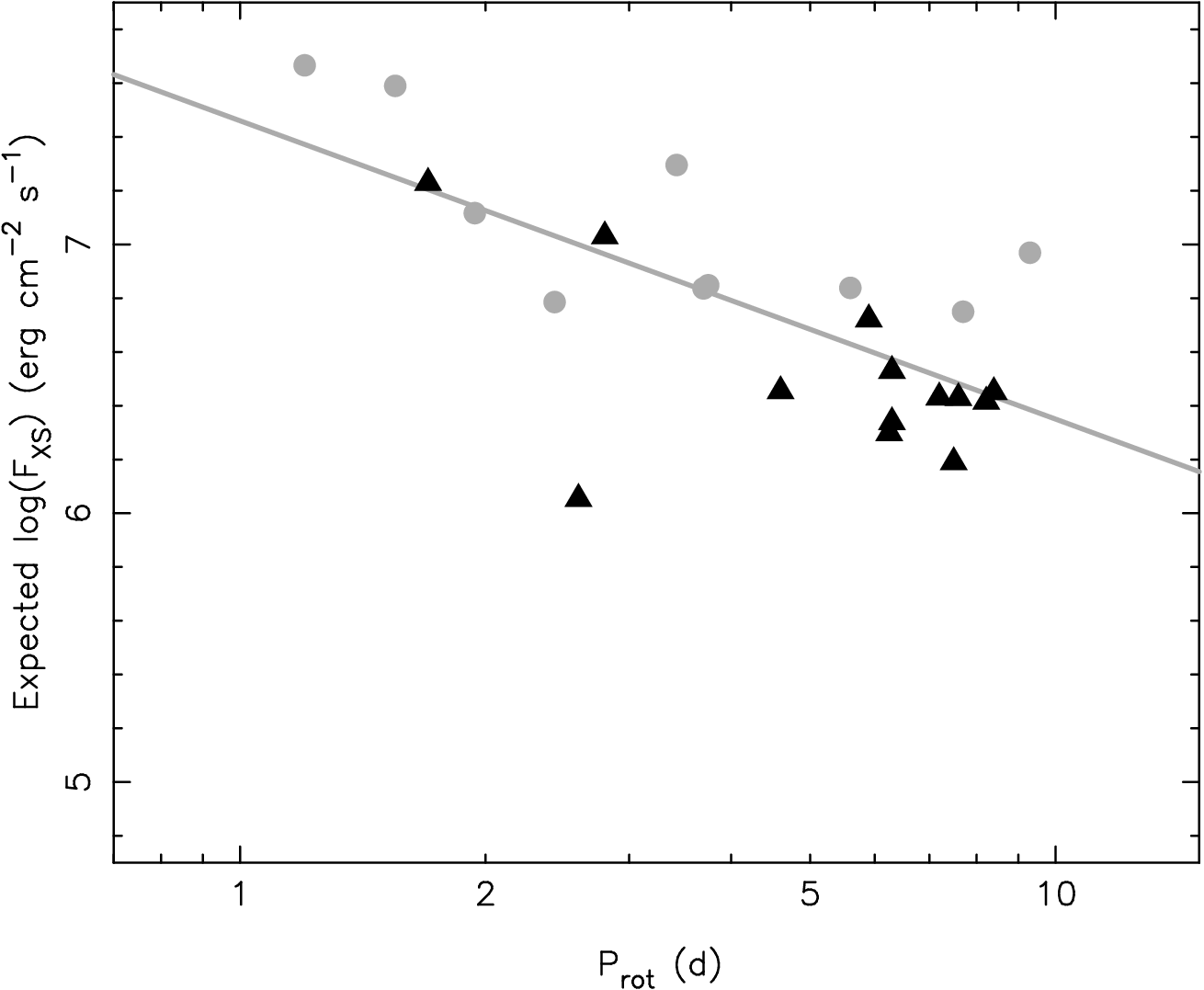}
}
\caption{The X-ray activity--rotation relation of T~Tauri stars in Taurus-Auriga observed by {\it XMM-Newton}. The observed (left) and expected (right) values of X-ray activity measures are plotted as a function of rotation period: Top, X-ray luminosity; middle, its ratio with stellar luminosity; bottom, the surface-averaged X-ray flux. Expected values were calculated for each star based on its stellar luminosity and whether it is accreting or not. Black triangles indicate accretors and grey circles indicate non-accretors. Lines show the best-fitting power-law correlation in each case.
}
\label{fig-xray-prot}
\end{figure*}

\subsection{Statistical analysis}
\label{sec-stats}

We used assumed power-law correlations between the X-ray activity measures and rotation period, and we have used the ASURV survival analysis package \citep{asurv} to perform the regression. Although ASURV is designed to analyse data containing upper and/or lower limits \citep{asurv2} while our sample of T~Tauri stars with measured photometric rotation periods contained no non-detections, this enabled us to compare our results directly with those of previous studies \citep[e.g.][]{sn01,coup}. We used the parametric Estimation Maximization (EM) method for the regression and have quoted the full range of probabilities of no correlation calculated by the three tests in ASURV (Cox Proportional Hazard, Generalized Kendall's $\tau$ and Spearman's $\rho$).

We used the two-sample tests in ASURV to assess the probabilities that pairs of observed distributions of $L_{\rm X}/L_{\star}$ were drawn from the same distribution (one of the subsamples of stars with no rotational information includes an upper limit). We have quoted the full range of probabilities calculated by the five tests in ASURV (Gehan's Generalized Wilcoxon Tests using permutation variance and hypergeometric variance, Logrank Test, Peto and Peto Generalized Wilcoxon Test, and Peto and Prentice Generalized Wilcoxon Test). 


\section{The observed activity--rotation relations and their origins}
\label{sec-act-rot}

\begin{table*}
\centering
\caption{Regression parameters and their standard deviations for linear fits of observed and expected values of $\log L_{\rm X}$, $\log(L_{\rm X}/L_{\star})$ and $\log F_{\rm XS}$ as a function of $\log P_{\rm rot}$. $A$ is the intercept and $B$ is the slope. The expected activity measures were calculated for each star based on its stellar luminosity and whether it was an accretor or not. The regression was performed using the Parametric Estimation Maximization method in the ASURV analysis package \citep{asurv}.}
\begin{tabular}{lcccccc}
\hline
Values & \multicolumn{2}{c}{$\log L_{\rm X}$ (erg\,s$^{-1}$)} & \multicolumn{2}{c}{$\log(L_{\rm X}/L_{\star})$} & \multicolumn{2}{c}{$\log F_{\rm XS}$ (erg\,cm$^{-2}$\,s$^{-1}$)}\\
 & $A$ & $B$ & $A$ & $B$ & $A$ & $B$\\
\hline
Observed & $31.04 \pm 0.21$ & $-1.20 \pm 0.30$ & $-3.20 \pm 0.16$ & $-0.37 \pm 0.24$ & $7.53\pm 0.17$ & $-1.04 \pm 0.25$\\
Expected & $30.98 \pm 0.25$ & $-1.27 \pm 0.36$ & $-3.27 \pm 0.10$ & $-0.44 \pm 0.15$ & $7.46 \pm 0.17$ & $-1.11 \pm 0.24$\\ 
\hline
\end{tabular}
\label{tab-xray-prot}
\end{table*}

Fig.~\ref{fig-xray-prot} (left) presents the relationship of the observed activity measures $L_{\rm X}$, $L_{\rm X}/L_{\star}$ and $F_{\rm XS}$ with rotation period for the 23 stars in our sample with measured photometric rotation periods. 
If we consider accretors and non-accretors together as a single sample, as has been done in previous studies, we observe clear anticorrelations of $L_{\rm X}$ and $F_{\rm XS}$ with rotation period. The tests performed in ASURV give probabilities of less than 0.01 that no correlation exists. The slopes of the correlations are approximately $-1$ (Table~\ref{tab-xray-prot}). However, we see no convincing anticorrelation of $L_{\rm X}/L_{\star}$ with rotation period; the tests in ASURV give probabilities of 0.14--0.26 that there is no correlation and the best-fit slope of approximately $-0.4$ is less than 1.5 standard deviations from zero (Table~\ref{tab-xray-prot}).

\citet{xest-acc} found a near-linear correlation of $L_{\rm X}$ with $L_{\star}$ in the XEST sample which they proposed was due to saturated activity. When they analysed the correlations of accretors and non-accretors separately, using the parametric EM algorithm in ASURV, they found similar slopes of $L_{\rm X}$ with $L_{\star}$ for the two samples but that $L_{\rm X}$ was systematically 0.4 dex (a factor 2.5) lower for accretors (see Fig.~\ref{fig-lx-ls}). The best-fitting relations were: $\log L_{\rm X} = (30.24 \pm 0.06) + (1.17 \pm 0.09) \log (L_{\star}/L_{\odot})$ for non-accretors and $\log L_{\rm X} = (29.83 \pm 0.06) + (1.16 \pm 0.09) \log (L_{\star}/L_{\odot})$ for accretors.

There is the impression in Fig.~\ref{fig-xray-prot} (left middle) that the accretors have consistently lower $L_{\rm X}/L_{\star}$ at all rotation periods, while the absence of a significant anticorrelation of $L_{\rm X}/L_{\star}$ with rotation period supports the concept of saturated activity. Nevertheless, the significant anticorrelations of $L_{\rm X}$ and $F_{\rm XS}$ with rotation period still suggest some kind of X-ray activity--rotation relation.

Let us assume that the X-ray activity of T~Tauri stars in Taurus-Auriga is saturated in the sense that $L_{\rm X}$, $L_{\rm X}/L_{\star}$ and $F_{\rm XS}$ have no \emph{intrinsic} dependence on rotation period, but $L_{\rm X}$ is instead characterized solely by the dependences on $L_{\star}$ and accretion reported by \citet{xest-acc}. We have thus calculated the expected value of $L_{\rm X}$, and hence $L_{\rm X}/L_{\star}$ and $F_{\rm XS}$, for each star in our sample based on its stellar luminosity and whether it is an accretor or non-accretor. Then we analysed the resulting correlation between each activity measure and rotation period as we did for the observed activity measures (Fig.~\ref{fig-xray-prot}, right). We find that significant anticorrelations of $L_{\rm X}$ and $F_{\rm XS}$, with probabilities of less than 0.01 that no correlation exists, result, along with a shallow anticorrelation of $L_{\rm X}/L_{\star}$. The intercepts and slopes of these anticorrelations are consistent with those found in the observed data (see Table~\ref{tab-xray-prot}). 

\begin{figure}
\centering
\includegraphics[angle=270,width=0.45\textwidth]{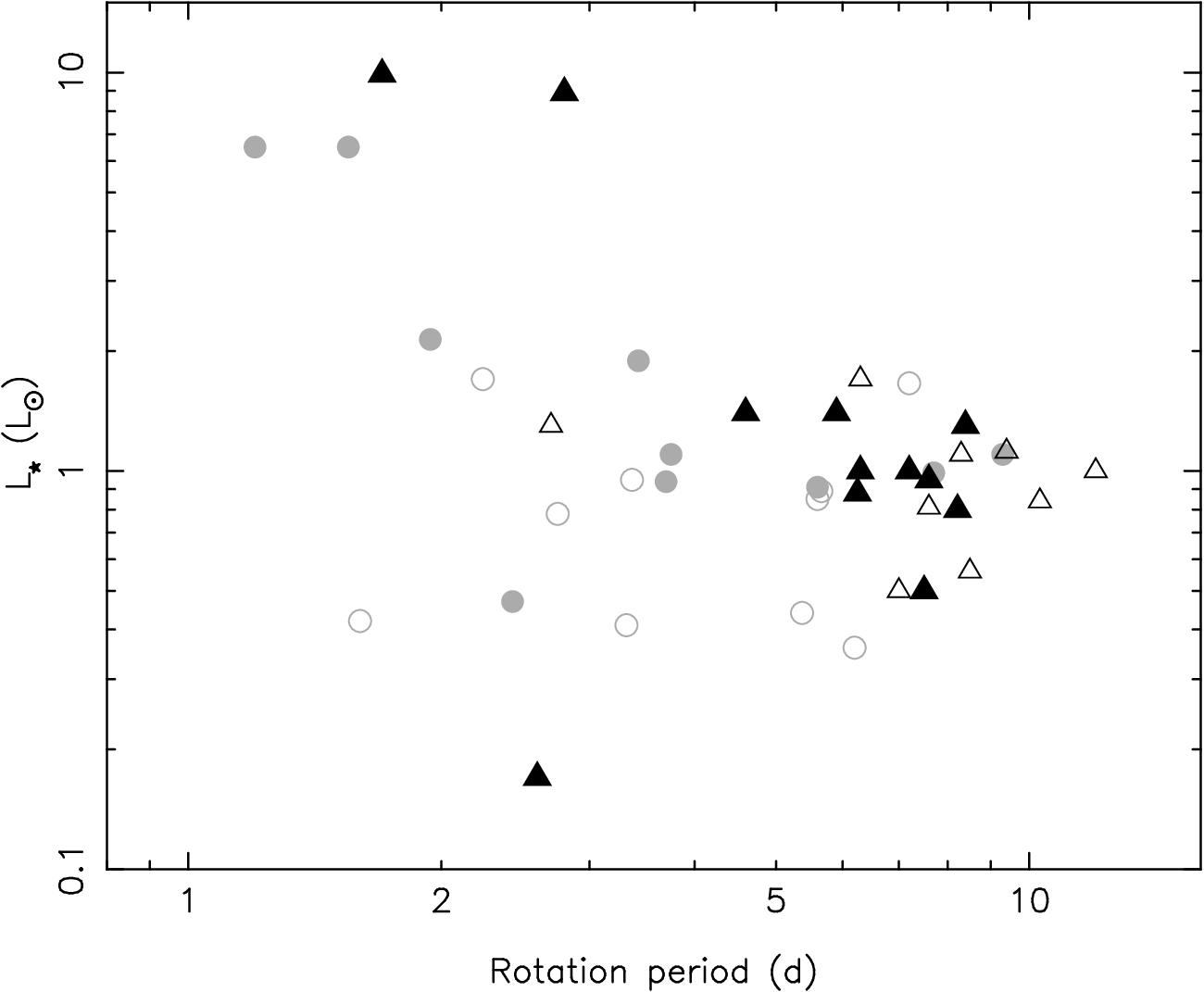}
\includegraphics[angle=270,width=0.45\textwidth]{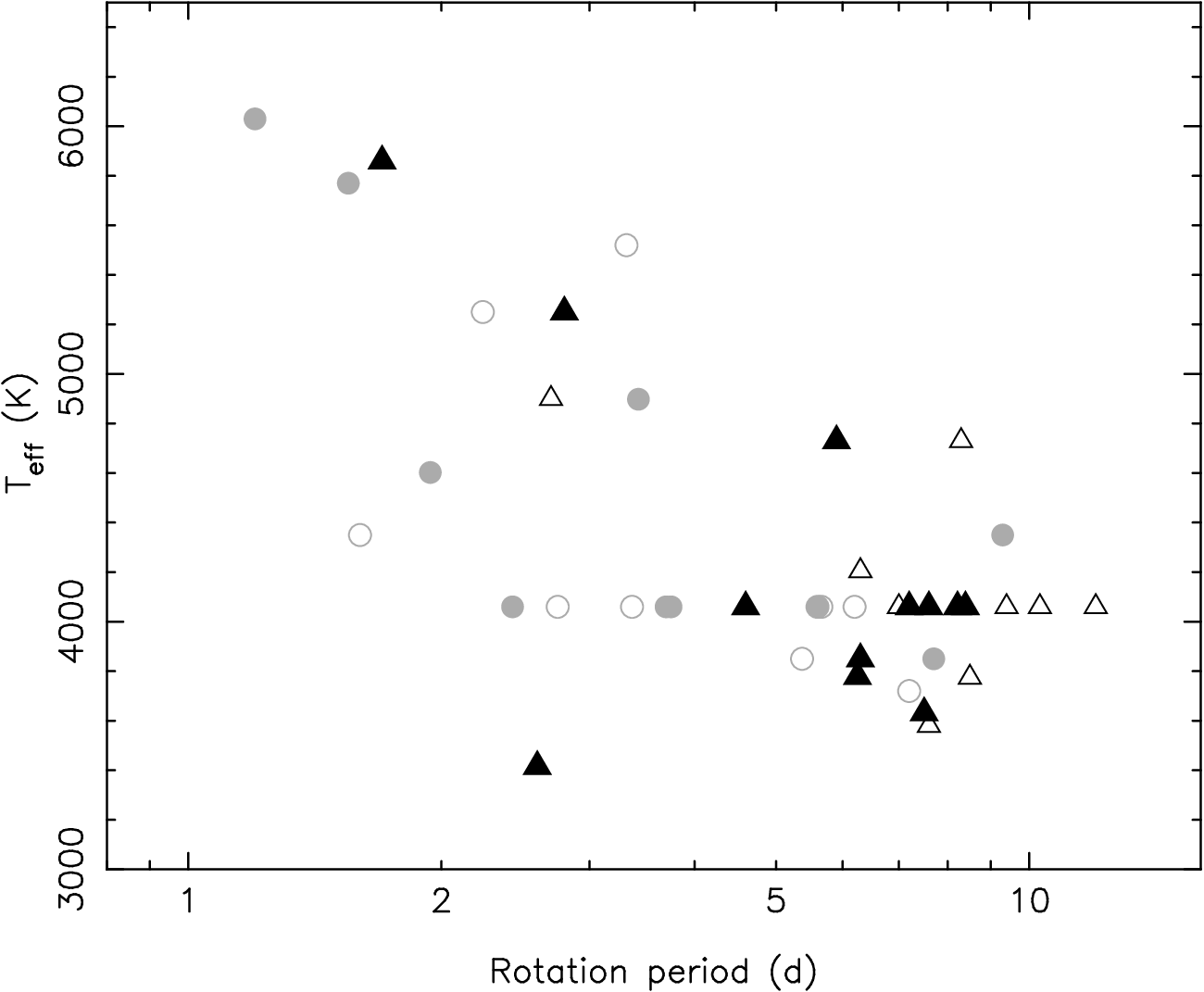}
\caption{The stellar luminosities (top) and effective temperatures (bottom) of T~Tauri stars in Taurus-Auriga with measured photometric rotation periods. Black triangles indicate accretors and grey circles mark non-accretors. Filled symbols denote stars observed by {\it XMM-Newton} in the present survey; unfilled symbols denote additional stars observed by {\it ROSAT} \citep{sn01}.}
\label{fig-ls-prot}
\end{figure}

The anticorrelation of $L_{\rm X}$ with rotation period can be explained by the the dependence of $L_{\rm X}$ on $L_{\star}$ and the observation that the fast rotators in Taurus-Auriga are typically more luminous, while the slow rotators are typically less luminous (Fig.~\ref{fig-ls-prot}, top). The shallow anticorrelation of $L_{\rm X}/L_{\star}$ with rotation period can be attributed to the lower $L_{\rm X}$ of accretors at any given $L_{\star}$ and the observation that the fast rotators in Taurus-Auriga are mainly non-accretors, while the slow-rotators are mainly accretors (Fig.~\ref{fig-ls-prot}). The anticorrelation of $F_{\rm XS}$ with rotation period then arises from the shallow $L_{\rm X}/L_{\star}$ relation due to the observation that the fast rotators in Taurus-Auriga have generally earlier spectral types, hence higher $T_{\rm eff}$ (Fig.~\ref{fig-ls-prot}, bottom), than the slow-rotators; recall $F_{\rm XS} = (L_{\rm X}/L_{\star}) \sigma T_{\rm eff}^4$. 

A plausible physical explanation for the dependence of rotation period on accretion and stellar luminosities and effective temperatures has been described by \citet{bou93}. T~Tauri stars spin up if they conserve angular momentum as they contract toward the main-sequence. Accreting stars may be prevented from spinning up either by losing angular momentum through their strong outflows or by being magnetically coupled to the rotation of their accretion disks, whereas non-accreting stars have been allowed to spin up as expected. More massive stars, which have higher $L_{\star}$ and $T_{\rm eff}$, may originate with faster rotation or may spin up more quickly due to their faster evolution toward the main-sequence. We note, however, that the lowest-mass T~Tauri stars (spectral types later than M2), which are largely excluded from our study, also appear to be typically fast rotators \citep{ppv-rot}, and other star-forming regions have different mass and age distributions from Taurus-Auriga, so the dependence of rotation on $L_{\star}$ and $T_{\rm eff}$, and hence the observed correlations of $L_{\rm X}$, $L_{\rm X}/L_{\star}$ and $F_{\rm XS}$ cannot be expected to be the same in all populations of T~Tauri stars.

An intrinsic dependence of X-ray activity on rotation period \emph{is not required} to produce the anticorrelations of $L_{\rm X}$ and $F_{\rm XS}$ on rotation period observed in Taurus-Auriga, and as such anticorrelations have not been reported in other populations such as the ONC, we consider there to be no convincing  evidence that they are intrinsic properties of the magnetic activity of T~Tauri stars.


\section{Origin of the lower $L_{\rm X}/L_{\star}$ of accretors}
\label{sec-lxls}
The lower $L_{\rm X}/L_{\star}$ of accretors compared to non-accretors in Taurus-Auriga has been reported before \citep{dam95,neu95,sn01}, and has usually been attributed to their slower rotation and the action of an activity--rotation relationship. However, the assumption we made in the previous section was that accretors had an intrinsically lower $L_{\rm X}/L_{\star}$ than non-accretors at all rotation periods.

If we analyse the correlation of $L_{\rm X}/L_{\star}$ with rotation period separately for the accretors and non-accretors, these samples, 13 accretors and 10 non-accretors, are very small and the resultant fits in ASURV have large uncertainties and low significances and can be strongly influenced by a single data point. The probability that no correlation exists is approximately 0.5 for both samples. The slope of $+0.16 \pm 0.19$ for the non-accreting sample suggests that $L_{\rm X}/L_{\star}$ is not anticorrelated with rotation period, but the slope of $-0.38 \pm 0.34$ for the accreting sample leaves open the possibility that rotation could play a role in the lower $L_{\rm X}/L_{\star}$ of accretors. The exclusion of \object{XZ~Tau}\footnote{\object{XZ~Tau} has the highest $L_{\rm X}/L_{\star}$ of the accretors, and has the lowest mass and stellar luminosity of any star in our sample with a measured rotation period. It is moreover a binary system whose secondary was known to be undergoing an optical outburst that influenced the X-ray emission of the system at the time of the observation used in XEST \citep{l1551b}.} reduces the slope for the accretors to $-0.11 \pm 0.34$.  

\begin{figure}
\centering
\includegraphics[angle=270,width=0.45\textwidth]{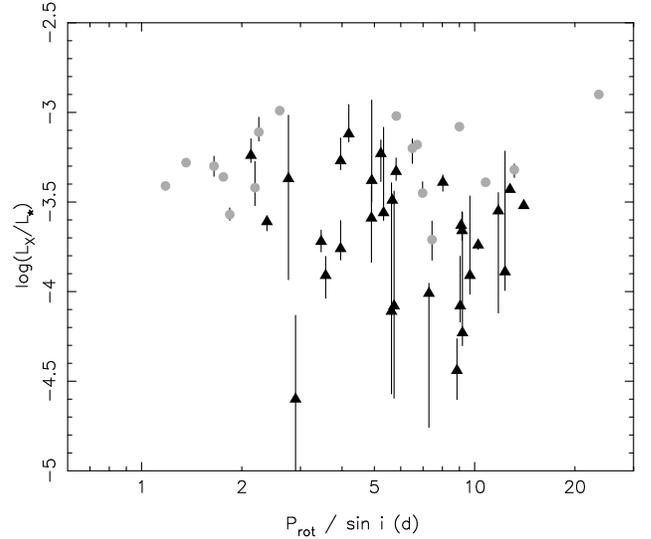}
\caption{The ratio of X-ray and stellar luminosities plotted as a function of rotation period for T~Tauri stars in Taurus-Auriga observed by {\it XMM-Newton} with measured $v \sin i$. Rotation periods have been calculated from $v \sin i$ and the stellar radius and the plotted values include an unknown factor $\sin i$ and are therefore upper limits. Triangles denote accretors and circles mark non-accretors.}
\label{fig-lxls-vsini}
\end{figure}

We have used the larger sample of T~Tauri stars with measured projected equatorial rotational velocities, $v \sin i$, to study in more detail the dependences of $L_{\rm X}/L_{\star}$ on rotation for accretors and non-accretors. 
This sample contains 30 accretors and 17 non-accretors. We have calculated rotation periods as $P_{\rm rot}/ \sin i = 2 \pi R_{\star} / v \sin i$. Fig.~\ref{fig-lxls-vsini} shows $L_{\rm X}/L_{\star}$ plotted against $P_{\rm rot}/ \sin i$. 


\begin{figure}
\centering
\includegraphics[angle=270,width=0.45\textwidth]{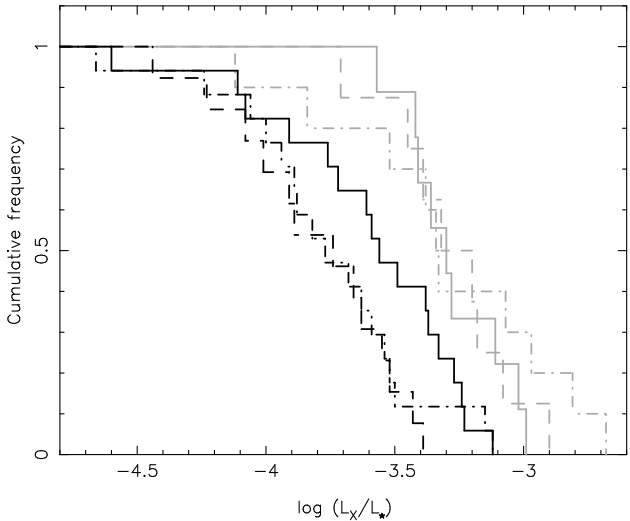}
\caption{Cumulative frequency distributions of $L_{\rm X}/L_{\star}$ for subsamples of T~Tauri stars in Taurus-Auriga observed by {\it XMM-Newton}. Grey lines show non-accretors and black lines show accretors. Unbroken lines mark fast rotators ($P_{\rm rot} / \sin i < 6$~d), dashed lines mark slow rotators ($P_{\rm rot} / \sin i > 6$~d), and dot-dashed lines mark stars with no measured $v \sin i$.}
\label{fig-two-sample}
\end{figure}

We have divided both the accretors and non-accretors into three subsamples: fast-rotators, with $P_{\rm rot}/ \sin i < 6$~d, slow rotators, with $P_{\rm rot}/ \sin i > 6$~d, and stars with no measured $v \sin i$. The cumulative frequency distributions of $\log (L_{\rm X}/L_{\star})$ for these subsamples are compared in Fig.~\ref{fig-two-sample}. We can immediately see that the distributions for the three subsamples of accretors are at lower $\log (L_{\rm X}/L_{\star})$ than the distributions for the three subsamples of non-accretors, and that the distribution for fast-rotating accretors is at higher $\log (L_{\rm X}/L_{\star})$ than those for the slow-rotating accretors. The mean $\log (L_{\rm X}/L_{\star})$ of each sample is given in Table~\ref{tab-two-sample}. The means of the three subsamples of non-accretors are well within $1\sigma$ of each other, while that of the fast-rotating accretors is approximately $2\sigma$ higher than those of the other accretors. For each of the three subsamples, the accretors have a mean $\log (L_{\rm X}/L_{\star})$ at least $3\sigma$ lower than the non-accretors. 

We have used the two-sample tests in ASURV to assess the probability, $P_{\rm same}$, that the distributions of $L_{\rm X}/L_{\star}$ in pairs of these subsamples were drawn from the same distribution. The distributions of $L_{\rm X}/L_{\star}$ for the subsamples of non-accretors are indistinguishable from one another ($P_{\rm same} > 0.5$). The distributions of $L_{\rm X}/L_{\star}$ for the  accretors which are slow rotators and those with no measured $v \sin i$ are also indistinguishable from one another ($P_{\rm same} > 0.75$), and not significantly different from that of the fast-rotating accretors ($P_{\rm same} = 0.09$--0.35 for slow-rotators and 0.16--0.24 for stars with no measured $v \sin i$). We find no strong evidence for a dependence of $L_{\rm X}/L_{\star}$ on rotation for accreting or non-accreting T~Tauri stars.

However, for each of the three subsamples, the distributions of $L_{\rm X}/L_{\star}$ of accretors and non-accretors are significantly different ($P_{\rm same} = 0.01$--0.03 for fast-rotators, $< 0.002$ for slow-rotators, and 0.005--0.03 for stars with no measured $v \sin i$). We therefore find good evidence for a property of accretors other than their slow rotation to be the reason for their lower $L_{\rm X}/L_{\star}$.

If we separate the samples at $P_{\rm rot}/ \sin i = 5$~d instead of 6~d, the distributions of $L_{\rm X}/L_{\star}$ of fast-rotating accretors and non-accretors are less significantly different ($P_{\rm same} = 0.03$--0.10) but the distributions of $L_{\rm X}/L_{\star}$ of accreting fast and slow rotators are even less significantly different ($P_{\rm same} = 0.18$--0.49). There is no support for a dependence of $L_{\rm X}/L_{\star}$ on rotation. 

The slightly higher $\log (L_{\rm X}/L_{\star})$ of fast-rotating accretors in our analysis may come about from an underestimation of $L_{\star}$ for some stars. $L_{\rm X}/L_{\star}$ and $P_{\rm rot}/ \sin i $ are not entirely independent variables as $P_{\rm rot} \sin i \propto R_{\star}$ and $R_{\star} \propto \sqrt{L_{\star}}$. If $L_{\star}$ is underestimated, a data point moves to the upper left in Fig.~\ref{fig-lxls-vsini}, to higher $\log (L_{\rm X}/L_{\star})$ and lower $P_{\rm rot}/ \sin i $. The $L_{\star}$ of accretors can be difficult to determine due to veiling and absorption. Three of the four accretors with $P_{\rm rot}/ \sin i < 5$~d and $\log (L_{\rm X}/L_{\star}) > -3.4$ have suspiciously high ages of $> 10$~Myr in the table of \citet{xest-overview} which suggest underestimated $L_{\star}$.

The accreting stars have noticeably larger error bars than the non-accretors. The accretors had typically lower-quality XEST X-ray spectra, partly due to their lower $L_{\rm X}$, which gives a lower X-ray flux from the star, and partly due to higher $N_{\rm H}$, which absorbs more of this X-ray flux. However, the upper limits of the larger error bars are produced by models with large amounts of very cool plasma ($T_0 \approx 2$~MK), absorbed by very high $N_{\rm H}$. As models with such cool temperatures were not found to fit the spectra of either accretors or non-accretors with low $N_{\rm H}$, we consider these unlikely to be realistic models. Such models would anyway point to fundamental differences in the coronae of these accretors and the non-accretors, in temperature rather than luminosity. 

We conclude that there is no demonstrable dependence of $L_{\rm X}/L_{\star}$ on rotation among T~Tauri stars in Taurus-Auriga. The activity is consistent with being saturated at $\log (L_{\rm X}/L_{\star}) \approx -3.3$ for accretors and $\approx -3.7$ for non-accretors. These values are consistent with those found for T~Tauri stars in the ONC \citep{coup}. The lower $L_{\rm X}/L_{\star}$ of accretors compared to non-accretors is not caused by their slower rotation, but to some other property that distinguishes accretors from non-accretors. The most natural interpretation is that it is a direct or indirect effect of the accretion process itself. We direct the reader to \citet{xest-acc} and \citet{coup} for discussions of mechanisms by which accretion could reduce the X-ray output of accreting stars.

Additional to these discussions, \citet{jardine} have proposed that the coronae of T~Tauri stars, and hence their X-ray emission, are limited by pressure stripping (resulting in lower-mass stars having lower $L_{\rm X}$, as observed e.g. in the ONC), but those of accreting T~Tauri stars may be further limited by the encroachment of the accretion disk. As a reasonable estimate for the location of the inner edge of the accretion disk is at the corotation radius, where orbiting material rotates around the star with the same period as the star itself rotates, Jardine et~al. further suggested that the wide range of $L_{\rm X}$ and $L_{\rm X}/L_{\star}$ observed among the accreting T~Tauri stars in the ONC could result from this rotation dependence. We have observed neither a wide range in $L_{\rm X}/L_{\star}$ among accretors nor a positive correlation of $L_{\rm X}$ or $L_{\rm X}/L_{\star}$ with rotation rate. However, a directly observable dependence on rotation could be swamped by other influential parameters in the model (e.g. stellar mass or coronal temperature), and the mass-dependence of $L_{\rm X}$ and the lower $L_{\rm X}/L_{\star}$ of accretors in the XEST sample reflect those observed in the ONC \citep{xest-acc}. Therefore, the Jardine et~al. model appears to be a viable description of the X-ray characteristics of T~Tauri stars that merits further observational testing.

\begin{table}
\centering
\caption{A comparison of the mean $\log(L_{\rm X}/L_{\star})$ of subsamples of T~Tauri stars in Taurus-Auriga observed by {\it XMM-Newton}. Stars in the \emph{fast} samples have $P_{\rm rot}/ \sin i < 6$~d, those in the \emph{slow} sample have $P_{\rm rot}/ \sin i > 6$~d, and those in the \emph{none} sample have no measured $v \sin i$.}
\begin{tabular}{lrcrc}
\hline
Sample & \multicolumn{2}{c}{Non-accretors} & \multicolumn{2}{c}{Accretors} \\
 & $N$ & $\langle \log(L_{\rm X}/L_{\star}) \rangle$ & $N$ & $\langle \log(L_{\rm X}/L_{\star}) \rangle$\\
\hline
Fast & 9 & $-3.27 \pm 0.06$ & 17 & $-3.61 \pm 0.09$\\
Slow & 8 & $-3.28 \pm 0.08$ & 13 & $-3.81 \pm 0.09$\\
None & 10 & $-3.31 \pm 0.13$ & 17 & $-3.79 \pm 0.10$\\
\hline
\end{tabular}
\label{tab-two-sample}
\end{table}

\section{Comparison with the activity of main-sequence stars and that of T~Tauri stars in the ONC}
\label{sec-rossby}

\begin{figure*}
\centering
\includegraphics[angle=270,width=0.85\textwidth]{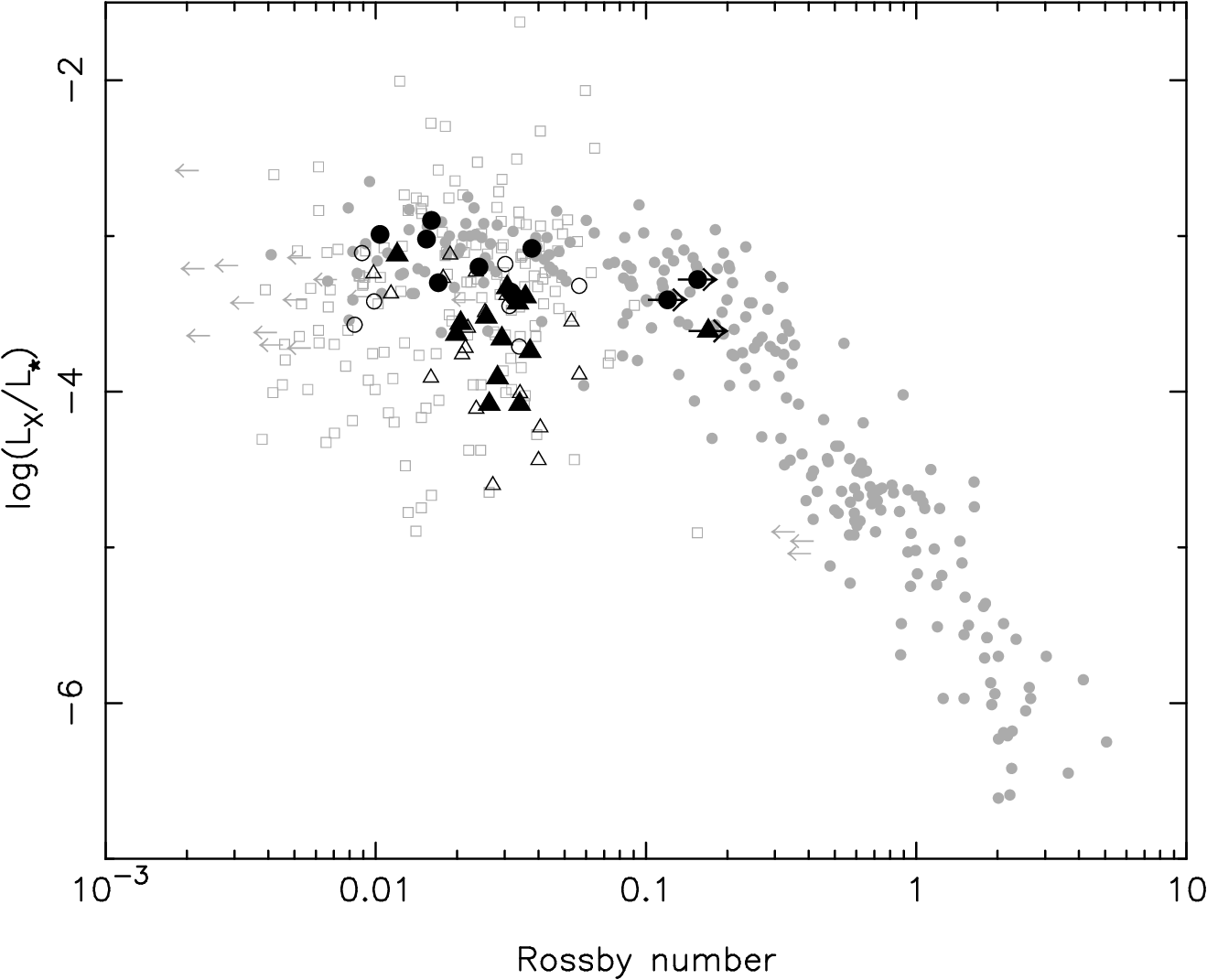}
\caption{The ratio of X-ray and stellar luminosities plotted as a function of Rossby number. Grey filled circles show late-type main sequence stars studied by \citet{pizzolato}; grey open squares show T~Tauri stars in the Orion Nebula Cluster studied by \citet{coup}; black circles and triangles show, respectively,  non-accreting and accreting T~Tauri stars in Taurus-Auriga in the present study. The filled black symbols show stars with measured photometric rotation periods; the open black symbols have periods derived from $v \sin i$ measurements and are effectively upper limits to the Rossby number. The grey leftward-pointing arrows represent periods derived from $v \sin i$ measurements in \citet{pizzolato}. The black rightward-pointing arrows show cases where only upper limits to the convective turnover time could be estimated.
}
\label{fig-rossby}
\end{figure*}

Fig.~\ref{fig-rossby} shows $L_{\rm X}/L_{\star}$ plotted as a function of Rossby number for T~Tauri stars in the XEST sample with measured photometric rotation periods, or for which rotation periods could be estimated from $v \sin i$ as in Sect.~\ref{sec-lxls}. Fig.~\ref{fig-rossby} also compares this sample with main-sequence stars studied by \citet{pizzolato} and T~Tauri stars in the COUP study of the ONC \citep{coup}. 

The K- and M-type T~Tauri stars in our sample all have values of $R_0$ that place them deep inside the saturated regime. The non-accretors with measured photometric rotation periods have $L_{\rm X}/L_{\star}$ and $R_0$ consistent with the main body of saturated main-sequence stars. The non-accretors with rotation periods derived from $v \sin i$ have a little lower $L_{\rm X}/L_{\star}$ but still within the range of values observed for saturated main-sequence stars. We conclude that the X-ray activity of non-accreting T~Tauri stars in Taurus-Auriga is entirely consistent with that of main-sequence stars showing saturated activity.

The accretors have Rossby numbers concentrated toward the higher end of the range of values of the non-accretors, due to their typically longer rotation periods, but well within the range of $R_0$ that defines saturated main-sequence stars. However, as we saw in Sect.~\ref{sec-lxls}, the $L_{\rm X}/L_{\star}$ of accretors is significantly lower than that of the non-accretors and a significant fraction have $\log (L_{\rm X}/L_{\star}) < -4$. Such an activity level would be clearly recognised as unsaturated for a late-type main-sequence star and interpreted as evidence for a less efficient magnetic dynamo due to slow rotation. If T~Tauri stars harbour the same kind of dynamo, the $\tau_{\rm conv}$ of these stars would need to have been systematically overestimated by approximately an order of magnitude for this interpretation to explain the lower activity of the accretors. \citet{wt03} have presented pre-main sequence evolutionary models in which T~Tauri stars that would be expected to be fully convective instead have radiative cores and a solar-like structure due to ongoing accretion. This would result in shorter convective turnover timescales than we have determined here, but detailed modelling would be required to determine how much shorter. Even if this would be the case, it would be the effect of accretion on the $\tau_{\rm conv}$ that would cause the lower X-ray activity of accretors compared to non-accretors, and not their slower rotation. Our results, however, point less to a reduced magnetic dynamo efficiency and more to the effects of circumstellar material on the stellar surface and atmosphere as an explanation of the lower X-ray emission of accretors \citep[see e.g.][]{xest-acc,coup}.


Also striking in Fig.~\ref{fig-rossby} are the positions of the G-type T~Tauri stars \object{SU~Aur}, \object{HD~283572} and \object{HP~Tau/G2}, for which we could estimate only upper limits to $\tau_{\rm conv}$. Although they are among the fastest rotators in our sample, their short $\tau_{\rm conv}$ moves them to the highest $R_0$, and they are almost outside the expected saturated regime. One might expect similar stars with longer rotation periods, perhaps $> 5$~d instead of 1--2~d, to fall outside the saturated regime and to show significantly lower $L_{\rm X}/L_{\star}$. This may have been observed among T~Tauri stars with radiative zones in NGC~2264 \citep{ngc2264}: while many had $L_{\rm X}/L_{\star}$ at saturated levels, a significant fraction also had $\log (L_{\rm X}/L_{\star}) < -5$. The mean $\log(L_{\rm X}/L_{\star})$ of the stars with radiative zones was significantly lower than that of those without. 
A population of stars with radiative zones showing such low activity is not found in the XEST sample. A two-sample test of stars with and without radiative zones finds no significant difference in the $\log L_{\rm X}/L_{\star}$ distributions of non-accretors or accretors ($P_{\rm same} = 0.15$--0.44 and 0.17--0.21, respectively). We propose that a significant proportion of the stars with radiative zones in NGC~2264 may be rotating slowly enough that their activity, produced in the same way as solar-like main-sequence stars, is no longer saturated.

The T~Tauri stars in Taurus-Auriga are concentrated to higher Rossby numbers than those in the ONC. This is mainly due to lower-mass T~Tauri stars, whose rotational properties are poorly studied in Taurus-Auriga, but which form a large population in the ONC and are mostly fast-rotating. The ONC stars have a larger scatter in $L_{\rm X}/L_{\star}$ (Fig.~\ref{fig-rossby}) and the ONC accretors and non-accretors have much more overlap in $L_{\rm X}/L_{\star}$ (not shown in Fig.~\ref{fig-rossby}) than those in Taurus-Auriga. This could indicate that younger T~Tauri stars experience greater X-ray variability, or it could reflect difficulties in estimating the $L_{\rm X}$ and $L_{\star}$ due to absorption. There are a number of non-accretors in the COUP data with $\log (L_{\rm X}/L_{\star}) < -4$ whose Rossby numbers indicate they should show saturated emission by analogy with solar-like main-sequence stars. Accretion cannot be the cause of lower activity in non-accreting T~Tauri stars, so such stars are potentially very interesting for investigating the differences between the X-ray activity of T~Tauri stars and late-type main-sequence stars.


\section{Comparison with previous results}
\label{sec-rosat}

The anticorrelations of $L_{\rm X}$, $F_{\rm XS}$ and $L_{\rm X}/L_{\star}$ with rotation period that we have found for the combined sample of accreting and non-accreting stars differ from those found by \citet{sn01} using {\it ROSAT} PSPC data. \citet{sn01} found steeper power-law slopes for all three activity measures, approximately $-1.5$ for $L_{\rm X}$ and $L_{\rm X}/L_{\star}$ and approximately $-2$ for $F_{\rm XS}$, and a highly significant anticorrelation of $L_{\rm X}/L_{\star}$ with rotation period.

We have made a careful reexamination of the data used by \citet[B.~Stelzer, priv. comm.]{sn01} to try to understand the differences between the two results. This has revealed systematic underestimation of $L_{\rm X}$ and overestimation of $L_{\star}$, both of which were stronger for accretors than for non-accretors. This caused a particularly strong underestimation of $L_{\rm X}/L_{\star}$, and hence $F_{\rm X}$, for some accretors with respect to non-accretors. Because the accretors have generally longer rotation periods than the non-accretors, this led to exaggerated slopes in the anticorrelations of the activity measures with rotation period. Significant correlations of $L_{\rm X}$ and $F_{\rm XS}$, and perhaps even $L_{\rm X}/L_{\star}$, with rotation period are nonetheless to be expected as the tendency for faster rotators to have higher $L_{\star}$ and $T_{\rm eff}$ and to be non-accreting is also true in the larger \citet{sn01} sample (Fig.~\ref{fig-ls-prot}). 

\begin{figure}
\centering
\includegraphics[angle=270,width=0.45\textwidth]{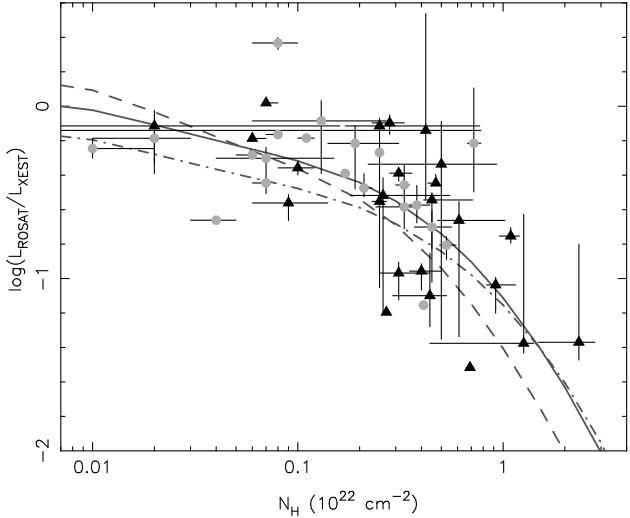}
\caption{Logarithm of the ratio of X-ray luminosities in the {\it ROSAT} study of \citet{sn01} and multiplicity-corrected X-ray luminosities in the present (XEST) study, plotted as a function of absorption column density, $N_{\rm H}$. Black triangles and grey circles mark accreting and non-accreting T Tauri stars, respectively. Lines show the expected ratio as a function of $N_{\rm H}$ for three spectral models covering the range of temperature considered in the XEST survey ($\log T_0 = 6.3$, dashed; 7.0, full; 7.5, dot-dashed, with $\beta = -1$ in each case), assuming a count-to-flux conversion factor of $1.0 \times 10^{-11}$\,erg\,cm$^{-2}$\,s$^{-1}$ per {\it ROSAT} PSPC count\,s$^{-1}$ was used to calculate the {\it ROSAT} luminosities. \citet{sn01} used this factor in cases where the hardness ratio could not be measured, for example, for the most-heavily absorbed stars which were not detected in the 0.1--0.4~keV band, while the maximum value used was $1.36 \times 10^{-11}$.}
\label{fig-rosat-ratio-nh}
\end{figure}

The underestimation of $L_{\rm X}$ was primarily due to insufficient accounting of the absorption of X-ray flux by interstellar and circumstellar material in the line of sight to the star. The {\it ROSAT} PSPC had, in principle, the spectroscopic capability to measure and account for the absorbing column density, $N_{\rm H}$. However, its soft energy bandpass of 0.1--2.4~keV and low sensitivity, especially at harder energies, combined with modest exposure times, provided spectra of insufficient quality for \citet{sn01} to perform spectral fitting for each star. They instead converted observed count-rates to X-ray fluxes using a conversion factor that aimed to account for absorption through a dependance on hardness ratio. Our calculations using the PSPC on-axis spectral response in XSPEC have found that the maximum value of this conversion factor would account for absorption only for $N_{\rm H} < 2.5 \times 10^{20}$~cm$^{-2}$, while almost all T~Tauri stars in Taurus-Auriga are observed through greater absorbing columns. Fig.~\ref{fig-rosat-ratio-nh} shows that 42 of the 45 stars detected in both the XEST and the {\it ROSAT} PSPC study have higher $L_{\rm X}$ in the XEST study\footnote{This includes also stars with spectral types later than M3.}, and that the factor of underestimation in \citet{sn01} is dependent on $N_{\rm H}$ in the manner expected and can exceed an order of magnitude. As accreting stars are generally surrounded by more circumstellar material than non-accretors, they are typically observed through higher $N_{\rm H}$ and therefore their $L_{\rm X}$ was typically underestimated by a larger factor. 

The overestimation of $L_{\star}$ in \citet{sn01} came about from the use of the bolometric luminosities listed by \citet{kh95}, as these include emission from circumstellar material (mostly in the infrared) as well as the output from the star itself. While the greater amount of circumstellar material around accretors led to a greater underestimation of $L_{\rm X}$ with respect to non-accretors, so it also led to a greater overestimation of their $L_{\star}$. 

A secondary reason for the underestimation of $L_{\rm X}/L_{\star}$ in \citet{sn01} was their treatment of multiple systems. The observed $L_{\rm X}$ was divided equally between the components, whereas we now know that $L_{\rm X}$ is approximately proportional to $L_{\star}$ so this approach typically underestimates the $L_{\rm X}$ of the primary component. At the same time, the bolometric luminosities from \citet{kh95} included emission from all components of multiple systems, and so overestimated the $L_{\star}$ of the primary. 

These difficulties demonstrate some of the challenges faced by studies of the X-ray activity of T~Tauri stars. While the higher-energy bandpasses and higher sensitivities of {\it XMM-Newton}'s EPIC and {\it Chandra}'s ACIS detectors have made it easier to account for absorption than was the case for the {\it ROSAT} PSPC, there are still cases where $N_{\rm H}$, and hence $L_{\rm X}$ are very poorly constrained. Uncertainties in $L_{\rm X}$ that account for uncertainties in the spectral fitting are rarely shown in the literature. Could the surprisingly large scatter in the COUP data be attributed to such uncertainties? 
The determination of the spectral type and extinction of accretors is complicated by veiling, the filling in of photospheric absorption lines by continuum emission from heated accreting material, which can make $L_{\star}$ also quite uncertain.
The division of the observed X-ray flux among the components of unresolved multiple systems is always somewhat arbitrary, and even stars which are not recognised to be multiples may yet have undiscovered companions. There remains room for improvement in studies of the X-ray emission of T~Tauri stars. While the usefulness of precision measurements is limited, because the $L_{\rm X}$ of any individual T~Tauri star is likely to be variable by factors of 2--3 on timescales of weeks--years and by larger factors during flaring on shorter timescales, it is important to avoid systematic problems, for example between accreting and non-accreting stars.


\section{Summary and Outlook}
\label{sec-summary}

We have used the {\it XMM-Newton} Extended Survey of the Taurus Molecular Cloud (XEST) to investigate the dependence of X-ray activity measures, $L_{\rm X}$, $L_{\rm X}/L_{\star}$ and $F_{\rm XS}$, on rotation period. 

Using only those stars with measured photometric rotation periods, we found power-law anticorrelations of $L_{\rm X}$ and $F_{\rm XS}$ with rotation periods that are significant at the 99 per cent level, and have slopes of approximately $-1$. However, we found no significant anticorrelation of $L_{\rm X}/L_{\star}$ with rotation period, with a best-fitting slope of approximately $-0.3$, which is consistent with saturation.

Although \citet{sn01} found steeper anticorrelations for all three of these activity measures, reexamination of the data used by \citet{sn01} has revealed that this can be explained by a systematic underestimation of $L_{\rm X}$ and overestimation of $L_{\star}$ that preferentially affected accreting stars, which are typically slow rotators.

\citet{xest-acc} have shown that the $L_{\rm X}$ of T~Tauri stars in Taurus-Auriga has a near-linear correlation with stellar luminosity, suggesting saturation, but that accreting stars have systematically lower $L_{\rm X}$ than non-accretors by a factor of 2.5 at any given $L_{\star}$. The anticorrelation of $L_{\rm X}$ with rotation period can be explained by the observation that the fast rotators in Taurus-Auriga have typically higher stellar luminosities than the slow rotators. The shallow anticorrelation of $L_{\rm X}/L_{\star}$ with rotation period can be attributed to the observation that the fast rotators in Taurus-Auriga are mainly non-accretors while the slow-rotators are mainly accretors. The anticorrelation of $F_{\rm XS} = (L_{\rm X}/L_{\star}) \sigma T_{\rm eff}^4$ with rotation period then comes about from the shallow $L_{\rm X}/L_{\star}$ relation due to the observation that the fast rotators in Taurus-Auriga have typically higher $T_{\rm eff}$ than the slow rotators.

Previous studies which have reported the lower $L_{\rm X}/L_{\star}$ of accretors in Taurus-Auriga have usually attributed it to the typically slower rotation periods of accretors and an anticorrelation of $L_{\rm X}/L_{\star}$ with rotation period. Using a larger sample of stars from the XEST sample with spectral types M3 or earlier, whose rotation periods were derived from $v \sin i$, we find no evidence that slow-rotating accretors have significantly lower $L_{\rm X}/L_{\star}$ than faster-rotating accretors, or that slow-rotating non-accretors have lower $L_{\rm X}/L_{\star}$ than fast-rotating non-accretors. This is consistent with saturated emission. However accretors were found to have significantly lower $L_{\rm X}/L_{\star}$ than non-accretors, whether they were fast-rotating, slow-rotating, or had no measured $v \sin i$.
 The lower $L_{\rm X}/L_{\star}$ of accretors compared to non-accretors is therefore not due to their slower rotation. 

\citet{pizzolato} have shown that the $L_{\rm X}/L_{\star}$ of solar-like main-sequence stars of spectral types G to early M is determined by the Rossby number, the ratio of rotation period to convective turnover timescale. We have estimated the convective turnover timescale of each star in the XEST sample by interpolating values calculated for T~Tauri stars in the ONC \citep{coup} and hence derived the Rossby number of each T~Tauri star in the XEST sample with rotational information. 

We have found that all T~Tauri stars of spectral types K and M in the XEST sample have Rossby numbers that lie well within the saturated regime of main-sequence stars. The non-accretors have $L_{\rm X}/L_{\star}$ entirely consistent with those of saturated main-sequence stars. The accretors have, however, generally lower $L_{\rm X}/L_{\star}$, and a significant fraction have $\log (L_{\rm X}/L_{\star}) < -4$. Although this would be interpreted as unsaturated emission if exhibited by a main-sequence star, the determined Rossby numbers of these accretors are approximately an order of magnitude too low to be in the conventional unsaturated regime. This points toward some effect of the coupling of the stellar surface and atmosphere to circumstellar material, rather than a reduced dynamo efficiency, as the cause of the lower X-ray output of accreting T~Tauri stars \citep[see][]{xest-acc,coup}. 

Identifying the process which causes this general reduction is an important motivation for future studies, but just one element in understanding the complex interaction of magnetic activity and the circumstellar environment. Recent studies have found, for instance, that dramatic increases in accretion rate can have quite differing effects on the observed X-ray emission in different cases \citep{v1647ori,v1118ori}, and that X-ray emission may also arise in shocks, in accretion streams onto the star \citep[e.g.][]{twhya} or in jet outflows \citep[e.g.][]{xest-jets}.

Non-accreting stars offer simpler means for understanding the magnetic activity and dynamo processes in pre-main sequence stars. The identification of non-accretors with $L_{\rm X}/L_{\star}$ significantly below the conventional saturation level and determination of their Rossby numbers is a potentially important probe of their dynamo mechanism. In particular, do there exist slow-rotating non-accreting T~Tauri stars with low activity that are analogous to the low-activity slow-rotating solar-like and fully-convective main-sequence stars and which would provide evidence for low dynamo efficiency due to slow rotation? 

 Investigations of the influence of stellar or circumstellar properties on the X-ray emission depend on a good characterization of those properties. The T~Tauri stars in Taurus-Auriga form a relatively small but very important population for such studies, because they are closeby and they and their circumstellar environments are individually well-studied. The small sample of just 23 stars with measured photometric rotation periods that we have studied here, out of a total of $\sim 200$ T~Tauri stars in Taurus-Auriga, demonstrates the potential that that further optical photometric monitoring and X-ray campaigns could unlock. However, there may remain still more potential in improved characterization of the stellar and circumstellar characteristics in the much larger, but more distant, sample of T~Tauri stars in the ONC in combination with refined analysis of the already comprehensive X-ray dataset obtained in the COUP campaign. It is also important to study pre-main sequence populations with a range of ages to understand the evolution of magnetic activity during this interesting phase of star and planet formation.


\begin{acknowledgements}
We thank the International Space Science Institut (ISSI) in Bern, Switzerland for logistic and financial support during several workshops on the XEST campaign. We are grateful for helpful discussions with other members of the XEST team during these workshops. We are also grateful to Y.-C.~Kim for his work in calculating convective turnover timescales for T~Tauri stars in the COUP. This research is based on observations obtained with {\it XMM-Newton}, an ESA science mission with instruments and contributions directly funded by ESA Member States and the USA (NASA). X-ray astronomy research at PSI has been supported by the Swiss National Science Foundation (grants 20-66875.01 and 20-109255/1). B.~S., L.~S., and G.~M. acknowledge financial contribution from contract ASI-INAF~I/023/05/0. M.~A. acknowledges support by NASA grant NNG05GF92G and from a Swiss National Science Foundation Professorship (PP002--110504).
\end{acknowledgements}


\end{document}